\documentclass[floatfix,footinbib,reprint,twocolumn,prb,a4paper,preprintnumbers,amsmath,amssymb]{revtex4-1}

\usepackage{graphics}

\usepackage{subfigure}
\usepackage{enumitem}  
\usepackage{bbold}
\DeclareGraphicsExtensions{.png,.pdf}
\usepackage{epsfig}
\usepackage{todonotes}
\usepackage{bm}
\usepackage{color}
\usepackage{makeidx}
\usepackage{hyperref}
	
\usepackage{booktabs}
\AtBeginDocument{
\heavyrulewidth=.08em
\lightrulewidth=.05em 
\cmidrulewidth=.01em
\belowrulesep=.65ex
\belowbottomsep=0pt
\aboverulesep=.4ex
\abovetopsep=0pt
\cmidrulesep=\doublerulesep
\cmidrulekern=.5em
\defaultaddspace=.5em
}
\usepackage{multirow}

\usepackage[utf8]{inputenc}
\usepackage{xcolor}

\begin{document}


\title{{\bf \textit{Ab-initio}} theory of the Gibbs free energy and a hierarchy of local moment correlation functions in itinerant electron systems: The magnetism of the Mn$_3$A materials class}

\author{Eduardo Mendive-Tapia}

\affiliation{Department of Physics, University of Warwick, Coventry CV4 7AL, U.K.}
\date{\today}

\author{Julie B. Staunton}

\affiliation{Department of Physics, University of Warwick, Coventry CV4 7AL, U.K.}
\date{\today}
\begin{abstract}

We present an \textit{ab-initio} disordered local moment theory for the Gibbs free energy of a magnetic material. Two central objects are calculated: the lattice Fourier transform of the direct local moment - local moment correlation functions in the paramagnetic state and local internal magnetic fields as functions of magnetic order. We identify the potentially most stable magnetic phases from the first, which can include non-collinear and long-period states in complex multi-atom unit cells, and extract higher order correlations among the local moments from the second. We propose that these latter entities produce a picture of effective  multi-site magnetic interactions depending on the state and extent of magnetic order and discuss its relation to other approaches.
We show how magnetic phase diagrams for temperature, magnetic field, and lattice structure and also magnetocaloric and mechanocaloric effects can be obtained from this approach.
The theory accurately predicts the order of transitions and quantifies contributions to first-order and order-order magnetic phase transitions from both purely electronic sources and magnetoelastic effects.
Our case study is the apparently frustrated magnetism of the Mn$_3$A class of materials in all its cubic, hexagonal, and tetragonal structures. The theory produces magnetic phases and transition temperatures in good agreement with experiment. We explain the first-order triangular antiferromagnetic to collinear antiferromagnetic transition in cubic Mn$_3$Pt as a magnetovolume driven effect, and its absence for A=Ir and Rh. We also construct the magnetic phase diagram of Mn$_3$Pt and explore its potential as a barocaloric material. Finally, we prepare the groundwork for future fully relativistic studies of the temperature dependence of the magnetism of Mn$_3$A, including Mn$_3$Sn, Mn$_3$Ga, and Mn$_3$Ge.

\end{abstract}


\maketitle

\section{Introduction}
\label{Intro}

We live in an age that is increasingly electronic and in which magnetic materials play a crucial role in technological applications.
Reliable description of temperature-dependent properties provided by first-principles calculations is necessary to explain fundamental aspects and predict behavior. For the majority of systems magnetic atoms host relatively long-lived local magnetic moments whose attributes become the appropriate magnetic degrees of freedom. Incorporating the effect of thermal fluctuations on these local moments is challenging and typically first-principles approaches are limited to the study of fully ordered magnetic states at zero temperature.
In this paper, however, we present a \textit{ab-initio} theory of magnetism which considers these thermally fluctuating magnetic degrees of freedom and which naturally describes the effect of magnetic exchange beyond pairwise interactions, magnetoelastic coupling, caloric effects for magnetic refrigeration, and which can identify the potential most stable magnetic phases.
Its task is to carefully describe how the electronic structure transforms in response to the formation of different ordered magnetic phases, as well as to lattice deformations, and how this produces a feedback on the evolution of the local moments themselves and their interactions.

Our theory's central outcome is an \textit{ab-initio} Gibbs free energy of a magnetic material,
\begin{widetext}
\begin{equation}
  \mathcal{G}_1= -T\sum_nS_n \big(\textbf{m}_n\big) +
 \left[\Omega_0+f^{(2)}\big(\{\textbf{m}_n\},\bm{\varepsilon}_{\alpha\beta}\big)+f^{(4)}\big(\{\textbf{m}_n\},\bm{\varepsilon}_{\alpha\beta}\big)+\dots\right]
-\textbf{H}\cdot\sum_n\mu_n\textbf{m}_
  +\frac{1}{2}C_{\alpha\beta\gamma\kappa}\bm{\varepsilon}_{\alpha\beta}\bm{\varepsilon}_{\gamma\kappa}+\sigma_{\alpha\beta}\bm{\varepsilon}_{\alpha\beta},
\label{EQ1}
\end{equation}
\end{widetext}
where we have split off a non-interacting part, the first term, from the rest (analogous to the density functional formalism developed for non-uniform classical fluids~\cite{doi:10.1080/00018737900101365}). This first term contains the entropy $\sum_nS_n$ from a mean field description. In the remainder $\Omega_0$ is a constant, $\textbf{H}$ is an external magnetic field, and $\{\mu_n\}$ are the sizes of local magnetic moments at atomic sites $\{n\}$, which fluctuate over time such that their orientations $\{\hat{e}_n\}$ are, on average, equal to local magnetic order parameters $\textbf{m}_n\equiv\langle\hat{e}_n\rangle$ (see Eq.\ (\ref{EQ5-Theo})). $\{f^{(a)}\}$ are order $a$ functions of $\{\textbf{m}_n\}$, i.e.\ $f^{(2)}$ describes pairwise correlations between local moments and $f^{(a>2)}$ higher order correlations among them.
These functions can also depend on any lattice deformation, quantifiable by a strain tensor $\bm{\varepsilon}_{\alpha\beta}$. The last two terms account for the cost of mechanical stress $\sigma_{\alpha\beta}$, described in terms of the fourth rank tensor $C_{\alpha\beta\gamma\kappa}$, a generalization of elastic moduli.
Crucially, $\{f^{(a)}\big(\{\textbf{m}_n\},\bm{\varepsilon}_{\alpha\beta}\big)\}$ are directly abstracted from first-principles calculations, using a Density Functional Theory (DFT) -based Disordered Local Moment (DLM) theory~\cite{0305-4608-15-6-018} which we have extended to make the following advances.
(1) For the first time we explain how magnetic correlations in the paramagnetic state can be used to study complex multi-atom per unit cell lattices and how they enable the potentially most stable complex magnetic structures to be found.
(2) We extend the theory to account for magnetoelastic effects such that the \textit{ab-initio} Gibbs free energy is minimized appropriately to include feedback between magnetic and structural degrees of freedom.
(3) Our theory can accurately find the order of magnetic phase transitions. We show how it provides two different sources for first-order and ordered-to-ordered magnetic transitions, one which has a purely electronic origin from the higher order local moment correlations and the other arising from magnetoelastic coupling.

The Mn$_3$A (A=Pt, Ir, Rh, Sn, Ga, Ge) class of magnetic materials in all its crystal structures (cubic, hexagonal, and tetragonal) is chosen as a striking case study. These systems are challenging tests for our theory owing to their large variety of phase transitions and magnetic phases, such as collinear and non-collinear antiferromagnetic states~\cite{NAGAMIYA1982385,doi:10.1143/JPSJ.51.2478,PhysRev.171.574,doi:10.1143/JPSJ.56.4532,doi:10.1063/1.2163485} and ferrimagnetism\cite{KREN19701653,Mn3GeINT}. The class has high relevance for spintronics~\cite{0295-5075-108-6-67001,0022-3727-47-30-305001,doi:10.1063/1.5021133,Zhange1600759,doi:10.1063/1.2722206,PhysRevB.83.020405,doi:10.1063/1.4970691,PhysRevLett.119.187204} and for the fundamental understanding of frustration~\cite{0953-8984-3-36-017,PhysRevB.93.184404} and spin chirality~\cite{PhysRevB.95.075128}.

The paper is organized as follows. We introduce our DFT-DLM theory in section \ref{MF}. A key part of our work is given in \ref{Wave}, where we use a linear response theory for the fully disordered paramagnetic state and obtain the magnetic pair correlations in this limit for complex multi-atom unit cells. This establishes a basis to explore potential magnetic phases that might form at lower temperatures.
We then show how to extract the full hierarchy of magnetic local moment correlations in section \ref{Multi}.
Section \ref{minim} explains how to describe magnetic properties as functions of temperature, magnetic field, and lattice structure~\cite{PhysRevLett.118.197202,PhysRevB.95.184438}.
The physics contained from the higher order local moment correlations is elaborated in section \ref{MSE}.
Section \ref{CE} is devoted to the calculation of caloric effects at finite temperature.
In section \ref{Mn3A} we present results for Mn$_3$A class of magnetic materials and in section \ref{Conc} we finally summarize the key theoretical aspects giving conclusions and outlook.

\section{\textit{Ab-initio} theory of magnetic ordering: disordered local moments and local magnetic fields}
\label{Theory}

In 1985 Gy\H{o}rffy \textit{et al}.\ \cite{0305-4608-15-6-018} established the conceptual framework from which Density Functional Theory (DFT) calculations can be used to describe the formation of local magnetic moments and their thermal fluctuations.
In the context of DFT, local moments
arise from the spin polarization of an electronic charge density.
Consider the magnetic moment formation process from spin-correlated electronic interactions on the fast time scale of electron propagation between atomic sites, $\tau_\text{elec}$. The central tenet is to assume that the orientations of these moments
are slowly varying degrees of freedom that remain constant over a time scale of $\tau_\text{form}>\tau_\text{elec}$. Time averages over $\tau_\text{form}$, therefore, render the system confined to a phase space prescribed by a collection of unitary vectors specifying the orientations of the local moments at each atomic site $n$, $\{\hat{e}_n\}$, i.e.\ the local spin polarization. The ergodicity is, in consequence, temporarily broken~\cite{doi:10.1080/00018738200101438,0305-4608-15-6-018}.
$\tau_\text{form}$ is also assumed to be short compared with the time $\mathcal{\tau_\text{wave}}$ necessary for the local moments to change their orientations, which has the scale of $\hbar/E_{SW}$, where $E_{SW}$ is a typical spin wave energy.
This picture is the basis of the Disordered Local Moment (DLM) theory~\cite{0305-4608-15-6-018}, which we use here. 
DFT-DLM theory has been successfully implemented in the past and recent years to study, for example, the onset of magnetic order in strongly-correlated systems~\cite{1367-2630-10-6-063010} and the heavy rare earth elements~\cite{Hughes1}, metamagnetic phase transitions in metal alloys~\cite{PhysRevB.87.060404,PhysRevB.89.054427}, the magnetic interactions between rare-earth and transition-metals~\cite{PhysRevLett.115.207201,PhysRevMaterials.1.024411}, and temperature-dependent magnetic anisotropy~\cite{PhysRevLett.93.257204,PhysRevB.74.144411,PhysRevB.90.054421,PhysRevLett.120.097202} and magnetostriction~\cite{PhysRevB.99.054415}.

Consider the grand potential functional of a magnetic material in the grand canonical ensemble~\cite{Fetter},
\begin{equation}
\Omega[\hat{\rho}]=\text{Tr}\hat{\rho}\left(\hat{\mathcal{H}}-\nu\hat{N}+\frac{1}{\beta}\log\hat{\rho}\right),
\label{EQ-T-DFT1}
\end{equation}
where $\nu$ is the chemical potential, $\hat{N}$ is the electron number operator, and $\hat{\mathcal{H}}$ is the electronic many-body Hamiltonian which can include an interaction with external fields.
Eq.\ (\ref{EQ-T-DFT1}) introduces the probability density operator $\hat{\rho}$ such that the expected value of an operator $\hat{A}$ is calculated as $\langle\hat{A}\rangle=\text{Tr}\,\hat{\rho}\hat{A}$, where $\text{Tr}\left[\dots\right]$ performs the trace operation over all degrees of freedom.
Systems whose magnetism is described well by the DLM picture enable $\text{Tr}$ to be approximated by the explicit separation of the trace over the orientations $\{\hat{e}_n\}$ and the remaining electronic degrees of freedom, that is~\cite{0305-4608-15-6-018}
\begin{equation}
\text{Tr}\rightarrow
\text{Tr}_{\{\hat{e}_n\}}\text{Tr}_\text{rest}.
\label{EQ1-Concep}
\end{equation}
By minimizing Eq.\ (\ref{EQ-T-DFT1}) with respect to the probability density and using Eq.\ (\ref{EQ1-Concep})
a grand potential constrained to a magnetic configuration $\{\hat{e}_n\}$ can be formally defined,
\begin{equation}
\Omega_c(\{\hat{e}_n\})=-\frac{1}{\beta}\log\left(\text{Tr}_\text{rest}\exp\left[-\beta(\hat{\mathcal{H}}-\nu\hat{N})\right]\right).
\label{EQ4-Concep}
\end{equation}
We emphasize that the directions $\{\hat{e}_n\}$ are classical quantities emerging from the many electron interacting system. In the context of DFT, this means that at short time scales $\tau_\text{form}$ the appropriate magnetization density, $\bm{\mu}(\textbf{r},\{\hat{e}_n\})$, is forced to satisfy
\begin{equation}
\mu_n(\{\hat{e}_n\})\hat{e}_n=\int_{V_n}{\text{d}\textbf{r}\,\bm{\mu}(\textbf{r},\{\hat{e}_n\})}.
\label{EQ5-concep}
\end{equation}
Eq.\ (\ref{EQ5-concep}) states that $\bm{\mu}(\textbf{r},\{\hat{e}_n\})$ is such that inside every region of volume $V_n$ centered at atomic site $n$ there is a total spin polarization, of size $\mu_n$, with an orientation constrained to be along $\hat{e}_n$
\footnote{Note that both $\bm{\mu}(\textbf{r},\{\hat{e}_n\})$ and the electron density $n(\textbf{r},\{\hat{e}_n\})$ are set to minimize a constrained grand potential functional
$\Omega_c[\hat{\rho}_\text{rest}]=\text{Tr}_\text{rest}\hat{\rho}_\text{rest}\left(\hat{\mathcal{H}}-\nu\hat{N}+\frac{1}{\beta}\log\hat{\rho}_\text{rest}\right)$,
i.e.\ for a probability density $\hat{\rho}_{\text{rest},0}=\frac{\exp\left[-\beta\left(\hat{\mathcal{H}}-\nu\hat{N}\right)\right]}{\text{Tr}_\text{rest}\exp\left[-\beta\left(\hat{\mathcal{H}}-\nu\hat{N}\right)\right]}$.}.

The equilibrium magnetic properties of the system are calculated by carrying out the ensemble averages over $\{\hat{e}_n\}$. Due to the continuous nature of $\{\hat{e}_n\}$, we apply $\text{Tr}_{\{\hat{e}_n\}}$ by performing the integrals $\text{Tr}_{\{\hat{e}_n\}}\rightarrow\prod_n\int{\text{d}\hat{e}_n}$.
Thus, in equilibrium every magnetic configuration is appropriately weighted by the probability $P(\{\hat{e}_n\})=\frac{\exp\left[-\beta\Omega_c(\{\hat{e}_n\})\right]}{\prod_{n'}\int{\text{d}\hat{e}_{n'}\,\exp\left[-\beta\Omega_c(\{\hat{e}_n\})\right]}}$.
The grand potential is, therefore, formally obtained by performing the average
\begin{equation}
\Omega=\prod_{n}\int{\text{d}\hat{e}_{n}P(\{\hat{e}_n\})
\left(\Omega_c(\{\hat{e}_n\})+\frac{1}{\beta}\log P(\{\hat{e}_n\})\right)}.
\label{EQ8-Concep}
\end{equation}
From the view of Eqs.\ (\ref{EQ4-Concep}) and (\ref{EQ8-Concep}), $\Omega_c(\{\hat{e}_n\})$ can be interpreted as a Hamiltonian governing the local moment orientations $\{\hat{e}_n\}$, which is characterized by the behavior of the rapidly responsive electronic structure and can be specified for every set $\{\hat{e}_n\}$.
The itinerant electron behavior makes the dependence on $\{\hat{e}_n\}$ of this object in principle very complicated in metals.

DFT-DLM theory has two conceptual pieces:
\begin{enumerate}
\item \textit{Evolution of the local moment orientations:} This part consists in performing the statistical mechanics of $\{\hat{e}_n\}$ to obtain the behavior of the interacting electron system at long time scales. Section \ref{MF} focuses on this task and establishes the framework to study the high temperature fully disordered regime in section \ref{Wave} and high order (beyond pair) local moment magnetic correlations in section \ref{Multi}.
\item \textit{Electronic structure part:} The second part refers to the description at short time scales $\tau_\text{form}$ and the formation of the local moments. The aim is to extract as best as possible the dependence of $\Omega_c(\{\hat{e}_n\})$ on the local moment directions required for the first part. The key theoretical quantities are therefore obtained from DFT calculations constrained to magnetic configurations $\{\hat{e}_n\}$. The Green's function based Multiple Scattering Theory (MST) formalism (Korringa-Kohn-Rostoker (KKR) method)~\cite{KORRINGA1947392,PhysRev.94.1111} is a pertinent technology for this task. In appendix \ref{S2APP} we give further details.
\end{enumerate}

\subsection{Mean-field theory and statistical mechanics of fluctuating local moments}
\label{MF}
$\Omega_c(\{\hat{e}_n\})$ describes the grand potential when the electron spin density $\bm{\mu}(\textbf{r},\{\hat{e}_n\})$ is constrained to have a specific orientational structure $\{\hat{e}_n\}$, one of huge number of non-collinear and collinear configurations.
It contains interactions among the local moments and their coupling with itinerant electron spin effects.
Contrary to magnetic insulators, in which the localized nature of the electrons and their spins usually allows for a description of their magnetism in terms of a simple Heisenberg type pairwise Hamiltonian, for metallic systems the magnetic interactions are in principle more complicated. As a consequence, $\Omega_c(\{\hat{e}_n\})$ might contain higher order magnetic interactions. This means that for example
\begin{equation}
\begin{aligned}
& \Omega_c(\dots
\uparrow\,
\uparrow\,
\uparrow\,
\uparrow\,
\uparrow\,
\uparrow\,
\uparrow
\dots)
+\Omega_c(\dots
\uparrow\,
\begingroup\color{red}\downarrow\endgroup\,
\uparrow\,
\uparrow\,
\uparrow\,
\begingroup\color{red}\downarrow\endgroup\,
\uparrow
\dots) \\
- & \Omega_c(\dots
\uparrow\,
\begingroup\color{red}\downarrow\endgroup\,
\uparrow\,
\uparrow\,
\uparrow\,
\uparrow\,
\uparrow
\dots)
-\Omega_c(\dots
\uparrow\,
\uparrow\,
\uparrow\,
\uparrow\,
\uparrow\,
\begingroup\color{red}\downarrow\endgroup\,
\uparrow
\dots),
\label{Cartoon1}
\end{aligned}
\end{equation}
which is a calculation designed to yield a two-site interaction, gives a different value than
\begin{equation}
\begin{aligned}
& \Omega_c(\dots
\downarrow\,
\uparrow\,
\downarrow\,
\uparrow\,
\downarrow\,
\uparrow\,
\downarrow
\dots)
+\Omega_c(\dots
\downarrow\,
\begingroup\color{red}\downarrow\endgroup\,
\downarrow\,
\uparrow\,
\downarrow\,
\begingroup\color{red}\downarrow\endgroup\,
\downarrow
\dots) \\
- & \Omega_c(\dots
\downarrow\,
\begingroup\color{red}\downarrow\endgroup\,
\downarrow\,
\uparrow\,
\downarrow\,
\uparrow\,
\downarrow
\dots)
-\Omega_c(\dots
\downarrow\,
\uparrow\,
\downarrow\,
\uparrow\,
\downarrow\,
\begingroup\color{red}\downarrow\endgroup\,
\downarrow
\dots).
\label{Cartoon1}
\end{aligned}
\end{equation}
Here arrows represent the local moment orientations for some simple collinear magnetic moment configuration.
The interactions between two sites inferred from calculations of the energy cost of rotating each of their local moment orientations through small angles, by for example using the Liechtenstein formula~\cite{LIECHTENSTEIN198765}, will depend upon the orientations of the moments on the surrounding sites. This is a direct consequence of these multi-site effects. Moreover, it is not tractable to evaluate  $\Omega_c(\{\hat{e}_n\})$ from the calculation of many non-collinear local moment configurations and the construction of large magnetic unit cells. Instead this paper shows how to capture these multi-site effects by focussing on the calculation of pair and higher order local moment correlation functions following ensemble averaging of $\Omega_c$. We construct a magnetic material's free energy and consequently follow their role in magnetic phase stabilization.

By working directly with $\Omega_c(\{\hat{e}_n\})$, the theory presented here introduces the following trial Hamiltonian as an efficient strategy to capture the evolution of $\{\hat{e}_n\}$,
\begin{equation}
\mathcal{H}_0(\{\hat{e}_n\})
=-\sum_n{\textbf{h}_n\cdot\hat{e}_n}.
\label{EQ1-Theo}
\end{equation}
Essentially, at each magnetic site the local moment experiences the effect of a local magnetic field $\textbf{h}_n$, set by the behavior of the many-interacting electrons and establishing a central mean-field approximation in our approach. $\textbf{h}_n$ can include an external applied magnetic field too if present.

The probability for a configuration $\{\hat{e}_n\}$ is
\begin{equation}
P_0(\{\hat{e}_n\})=\frac{1}{Z_0}\exp\left[-\beta\mathcal{H}_0(\{\hat{e}_n\})\right]=\prod_n{P_n(\hat{e}_n)},
\label{EQ2-Theo}
\end{equation}
where $P_n(\hat{e}_n)=\frac{\exp\left[\beta\textbf{h}_n\cdot\hat{e}_n\right]}{Z_{0,n}}$ are single-site probabilities,
which depend on the direction and length of $\boldsymbol{\lambda}_n\equiv\beta\textbf{h}_n$, and $Z_0 =\prod_n{Z_{0,n}}=\prod_n{\int{\text{d}\hat{e}_n\exp\left[\boldsymbol{\lambda}_n\cdot\hat{e}_n\right]}}=\prod_n{4\pi\frac{\sinh\lambda_n}{\lambda_n}}$ is the partition function,
We can use now Eq.\ (\ref{EQ2-Theo}) to carry out the average over $\{\hat{e}_n\}$. For example,
\begin{equation}
\left\{\textbf{m}_n=\int{\text{d}\hat{e}_nP_n(\hat{e}_n)}\hat{e}_n=\left(\frac{-1}{\lambda_n}+\coth\lambda_n\right)\hat{\lambda}_n\right\},
\label{EQ5-Theo}
\end{equation}
which describe the amount of magnetic order at every magnetic site associated with the orientational configurations of the local moments, i.e.\ they are magnetic local order parameters.
A magnetic phase is fully specified by a particular set $\{\textbf{m}_n\}$. For example, the paramagnetic state in which all magnetic sites are fully disordered corresponds to $\{\textbf{m}_n\}=\{\textbf{0}\}$, a ferromagnetic state to $\{\textbf{m}_n\}=\{\textbf{m}_\text{FM}\}$, and a helical antiferromagnetic order modulated by a wave vector $\textbf{q}_0=(0, 0, q_0)$ applies when $\textbf{m}_n=m_0\left[\cos(\textbf{q}_0\cdot\textbf{R}_n),\sin(\textbf{q}_0\cdot\textbf{R}_n),0)\right]$, where $\textbf{R}_n$ is the position of the $n$th site.

The Gibbs free energy associated with $\mathcal{H}_0$ is given by $\mathcal{G}_0=\frac{-1}{\beta}\log Z_0$.
Since $\mathcal{H}_0$ is a trial Hamiltonian, according to the Peierls-Feynman inequality~\cite{PhysRev.97.660,doi:10.1143/JPSJ.50.1854} an upper bound $\mathcal{G}_1$ of the exact Gibbs free energy $\mathcal{G}$, associated with the Hamiltonian $\Omega_c(\{\hat{e}_n\})$, is
\begin{equation}
\mathcal{G}_1=\mathcal{G}_0+\Big\langle\Omega_c(\{\hat{e}_n\})-\mathcal{H}_0(\{\hat{e}_n\})\Big\rangle_0\ge\mathcal{G},
\label{EQ7-Theo}
\end{equation}
where $\langle\cdots\rangle_0$ is the ensemble average with respect to the trial probability $P_0(\{\hat{e}_n\})$. Under the presence of an external magnetic field $\textbf{H}$, that couples with local moments with sizes $\{\mu_n\}$, we can write the average
\begin{equation}
\langle\Omega_c\rangle_0  =\Big\langle\Omega^\text{int}-\sum_n\mu_n\hat{e}_n\cdot\textbf{H}\Big\rangle_0 \\
  =\langle\Omega^\text{int}\rangle_0-\sum_n\mu_n\textbf{m}_n\cdot\textbf{H},
\label{EQ7a-Theo}
\end{equation}
where we have explicitly separated the term accounting for the coupling with $\textbf{H}$. After some simple algebra it follows that
\begin{equation}
\mathcal{G}_1=\langle\Omega^\text{int}\rangle_0-\sum_n\mu_n\textbf{m}_n\cdot\textbf{H}-TS_\text{mag}.
\label{EQ11-Theo}
\end{equation}
where $S_\text{mag}=\sum_n{S_n}(\textbf{m}_n)$
is the magnetic entropy contribution from the orientational local moment configurations, containing single-site magnetic entropies which appear in the first term of Eq.\ (\ref{EQ1}),
\begin{equation}
\begin{split}
 & S_n(\textbf{m}_n)=-k_\text{B}\int{\text{d}\hat{e}_n P_n(\hat{e}_n)\log P_n(\hat{e}_n)}  \\
 &               =k_\text{B}\left[1+\log\left(4\pi\frac{\sinh\lambda_n}{\lambda_n}\right)-\lambda_n\coth \lambda_n\right].
\label{EQ10-Theo}
\end{split}
\end{equation}
Since the average over the magnetic configurations is taken with respect to $P_0(\{\hat{e}_n\})$, the natural order parameters of $\mathcal{G}_1$ are $\{\textbf{m}_n\}$. To ensure that $\mathcal{G}_1$ is minimized with respect to them so that Eq.\ (\ref{EQ7-Theo}) is exploited we write
\begin{equation}
-\nabla_{\textbf{m}_n}\mathcal{G}_1=\textbf{h}^\text{int}_n+\mu_n\textbf{H}-\textbf{h}_n=\textbf{0},
\label{EQ12-Theo}
\end{equation}
where $\textbf{h}_n=\frac{\partial\left[-TS_\text{mag}\right]}{\partial\textbf{m}_n}$ follows from Eqs.\ (\ref{EQ5-Theo}) and (\ref{EQ10-Theo}).
Eq.\ (\ref{EQ12-Theo}) defines the internal, local magnetic fields
\begin{equation}
\left\{\textbf{h}^\text{int}_n=-\frac{\partial\langle\Omega^\text{int}\rangle_0}{\partial\textbf{m}_n}\right\},
\label{EQ14-Theo}
\end{equation}
and shows that the equilibrium condition is 
\begin{equation}
\left\{\textbf{h}_n=\textbf{h}_n^\text{int}+\mu_n\textbf{H}\right\}
.
\label{EQ15-Theo}
\end{equation}
Eqs.\ (\ref{EQ5-Theo}), (\ref{EQ14-Theo}), and (\ref{EQ15-Theo}) are the central equations describing the magnetic state in equilibrium.
From the perspective of Eqs.\ (\ref{EQ1-Theo}) and (\ref{EQ5-Theo}), $\textbf{h}_n$ should be regarded as the effective field at site $n$ that would sustain the local moment with an averaged orientation equal to $\textbf{m}_n$. 
Moreover, the physical meaning of $\textbf{h}_n^\text{int}$ is given by Eq.\ (\ref{EQ14-Theo}), which shows that it is the emerging local magnetic field when the electronic structure is forced to coexist with a magnetic ordering prescribed by $\{\textbf{m}_n\}$, as imposed by the averaging over $P_0(\{\hat{e}_n\})$. Hence, Eq.\ (\ref{EQ15-Theo}) has a clear physical interpretation: At equilibrium the total local magnetic field, composed by the addition of $\textbf{h}^\text{int}_n$ and $\textbf{H}$, must be identical at every site to the magnetic field necessary to sustain the local moments whose fluctuating magnetic orientations are, on average, $\{\textbf{m}_n\}$.

The Green's function based KKR-DFT formalism, together with the Coherent Potential Approximation (CPA) employed to carry out the averages over the magnetic configurations, is used to calculate the right hand side of Eq.\ (\ref{EQ14-Theo}), and so extract $\{\textbf{h}^\text{int}_n\}$ for a chosen input of $\{\textbf{m}_n\}$ (or equivalently for a chosen input of $\{P_n(\hat{e}_n)\}$, which in turn are prescribed by $\{\boldsymbol{\lambda}_n\}=\{\beta\textbf{h}_n\}$, as shown by Eqs.\ (\ref{EQ2-Theo}) and (\ref{EQ5-Theo}))~\cite{0305-4608-15-6-018}.
The theory is designed, therefore, to describe the dependence of the local magnetic fields on the state of magnetic order.
We finally point out that a simple classical Heisenberg model with constants $\{J_{ij}\}$ would map into a linear magnetic dependence described as $\textbf{h}_i^\text{int}(\{\textbf{m}_n\})=\sum_{j}J_{ij}\textbf{m}_{j}$ (see section \ref{Multi}). However, this simple form is not guaranteed for magnetic materials with a complicated dependence of the electronic structure on $\{\textbf{m}_n\}$, as for example proven in the past for many metallic magnetic materials~\cite{PhysRevLett.115.207201,PhysRevLett.118.197202,PhysRevB.95.184438,PhysRevX.8.041035}. The Mn$_3$A class of materials provides a rich framework to study this as we will show in section \ref{Mn3A}.

\subsection{Potential multi-atom per unit cell complex magnetic phases from fully disordered local moments}
\label{Wave}

We investigate the formation and stability of different magnetic phases by calculating the corresponding local fields $\{\textbf{h}^\text{int}_n\}$, emerging from the electronic structure, as functions of magnetic order $\{\textbf{m}_n\}$.
Studying all possible competing magnetic phases, which can include long-period and non-collinear states for example, is however a formidable challenge.
The strategy we follow is based on firstly analyzing the magnetic correlations in the fully disordered high temperature paramagnetic (PM) state ($\{\textbf{m}_n\}\rightarrow\{\textbf{0}\}$) and gain information on the potential magnetic ordered states the PM state might become unstable to at a lower temperature~\cite{0305-4608-15-6-018}. In general,
\begin{equation}
\textbf{h}^\text{int}_i=\sum_j \mathcal{S}^{(2)}_{ij}\textbf{m}_j+\text{higher order terms},
\label{EQ1a-Wave}
\end{equation}
which from Eq.\ (\ref{EQ14-Theo}) defines the so-called direct correlation function as the second derivative of $\langle\Omega^\text{int}\rangle_0$ with respect to the local order parameters in the PM limit, and which comprises the $f^{(2)}\big(\{\textbf{m}_n\},\bm{\varepsilon}_{\alpha\beta}\big)$ component in the square brackets of the second term in  Eq.\ (\ref{EQ1}) for the Gibbs free energy,
\begin{equation}
\mathcal{S}^{(2)}_{ij} = -\frac{\partial^2\langle\Omega^\text{int}\rangle_0(\{\textbf{m}_n\})}{\partial\textbf{m}_i\partial\textbf{m}_j}\Bigg|_{\{\textbf{m}_n\}=\{0\}}.
\label{EQ1-Wave}
\end{equation}
The magnetic correlations in the PM state are contained in $\mathcal{S}^{(2)}_{ij}$.
To examine them we study the effect of applying an infinitesimally small site-dependent external magnetic field $\textbf{H}_n$, which induces small magnetic polarizations $\{\delta\textbf{m}_n\}$ at each site $n$,
\begin{equation}
\delta\textbf{m}_i=\sum_j\chi_{ij}(\{\textbf{m}_n\})\textbf{H}_j,
\label{EQ2-Wave}
\end{equation}
where the linear response to the magnetic field application is described by the magnetic susceptibility $\chi_{ij}(\{\textbf{m}_n\})$.
Using Eq.\ (\ref{EQ1-Wave}) and recalling that $\textbf{h}_n=\textbf{h}^\text{int}_n+\textbf{H}_n$ in equilibrium (see Eq.\ \ref{EQ15-Theo}) we can write
\begin{equation}
\chi_{ij}=\frac{\beta}{3}\left(\delta_{ij}+\sum_{k}{\mathcal{S}^{(2)}_{ik}\chi_{kj}}\right),
\label{E6-Wave}
\end{equation}
where we have used that $\left\{\textbf{m}_n\approx\frac{\beta}{3}\textbf{h}_n\right\}$
in the PM limit, as directly follows from Eq.\ (\ref{EQ5-Theo}).
\begin{equation}
\zeta^\text{PM}_{ij}=\left(\chi^{-1}\right)_{ij}=\frac{\partial^2\mathcal{G}_1}{\partial\textbf{m}_i\partial\textbf{m}_j}\Bigg|_{\{\textbf{m}_n\}=\{0\}}=3k_\text{B}T\delta_{ij}-\mathcal{S}^{(2)}_{ij},
\label{EQ7aa-Wave}
\end{equation}
are the components of the Hessian matrix associated with the Gibbs free energy $\mathcal{G}_1$ introduced in Eq.\ (\ref{EQ11-Theo}) where the first term comes from the second derivative of the non-interacting term of
 Eq.\ (\ref{EQ1}) containing the local entropy.
The transition temperature $T_\text{max}$ below which the PM state is unstable to the formation of a magnetic phase can be calculated by solving the condition for the magnetic susceptibility to diverge, i.e.\ $\det\left(\zeta^\text{PM}\right)=0$,
which reduces to a calculation of the eigenvalues $\{u_a\}$ of $\mathcal{S}^{(2)}_{ij}$ which satisfy
\begin{equation}
\prod_{a=1}^{N_m}\big(3k_\text{B}T-u_a\big)=0,
\label{EQ7c-Wave}
\end{equation}
where $N_m$ is the number of magnetic atoms in the crystal. $T_\text{max}$is found from the largest eigenvalue $u_\text{max}$ of $\mathcal{S}^{(2)}_{ij}$,
\begin{equation}
T_\text{max}=\frac{u_\text{max}}{3k_\text{B}}.
\label{EQ7d-Wave}
\end{equation}

Solving the eigenvalue problem set by $\mathcal{S}^{(2)}_{ij}$ in the real space is in principle very complicated if there are long range magnetic correlations, requiring the diagonalization of a high dimensional matrix as well as the calculation of its components. We instead exploit the crystal symmetry in the PM state and apply a lattice Fourier transform defined as follows
\begin{equation}
\label{EQ7-Wave3}
\tilde{\mathcal{S}}^{(2)}_{ss'}(\textbf{q})=\frac{1}{N_\text{c}}\sum_{tt'}{\mathcal{S}^{(2)}_{ts\,t's'}\exp{\big[-i\textbf{q}\cdot\left(\textbf{R}_{t}-\textbf{R}_{t'}\right)}\big]}, 
\end{equation}
where $N_\text{c}$ is the number of unit cells. We have decomposed each lattice site index $i$ and $j$ into two additional indices, $i\rightarrow t,s$ and $j\rightarrow t',s'$. We use $t$ and $t'$ to specify the origin of unit cells, and $s$ and $s'$ for indices denoting the atomic positions within the unit cells $t$ and $t'$, respectively. $\textbf{R}_{t}+\textbf{r}_{s}$ is, therefore, the position vector of the magnetic atom at site $(t,s)$ such that $\textbf{R}_{t}$ denotes the origin of the unit cell $t$ and $\textbf{r}_{s}$ gives the relative position of the sub-lattice $s$ within that unit cell. $\tilde{\mathcal{S}}^{(2)}_{ss'}(\textbf{q})$ and $\tilde{\zeta}_{ss'}(\textbf{q})$ are square matrices whose components are functions of the wave vector $\textbf{q}$. Their dimension is the number of magnetic positions, or sub-lattices, inside the unit cell, $N_\text{at}=N_m/N_c$. Note that the shape and components of these matrices depend on the choice of the unit cell.
Lattice Fourier transforming Eq.\ (\ref{EQ7aa-Wave}) gives
\begin{equation}
\tilde{\zeta}^\text{PM}_{ss'}(\textbf{q})=3k_\text{B}T\delta_{ss'}-\tilde{\mathcal{S}}^{(2)}_{ss'}(\textbf{q}).
\label{EQ8-Wave2}
\end{equation}
Hence, by applying the lattice Fourier transform we have reduced the dimension of the matrix to be diagonalized from $N_m\times N_m$ to $N_\text{at}\times N_\text{at}$. Thus, Eq.\ (\ref{EQ7c-Wave}) becomes
\begin{equation}
\prod_{a=1}^{N_\text{at}}\big(3k_\text{B}T-\tilde{u}_a(\textbf{q})\big)=0,
\label{EQ10-Wave}
\end{equation}
where now $\{\tilde{u}_a(\textbf{q}),\,a=1,\,\dots,\,N_\text{at}\}$ are the $N_\text{at}$ eigenfunctions of $\tilde{\mathcal{S}}^{(2)}_{ss'}(\textbf{q})$, which depend on the wave vector $\textbf{q}$ and are obtained by solving the eigenvalue problem
\begin{equation}
\sum_{s'=1}^{N_\text{at}}\tilde{\mathcal{S}}^{(2)}_{ss'}(\textbf{q})V_{a,s'}(\textbf{q})=\tilde{u}_a(\textbf{q})V_{a,s}(\textbf{q}),
\label{EQ11-Wave}
\end{equation}
for $\{s=1,\,\dots,\,N_\text{at}\}$, and $N_\text{at}$ eigenvectors of dimension $N_\text{at}$, $\{(V_{a,1},\,\dots,\,V_{a,N_\text{at}}),\,a=1,\,\dots,\,N_\text{at}\}$.
From Eq.\ (\ref{EQ10-Wave}) we obtain
\begin{equation}
T_\text{max}=\frac{\tilde{u}_\text{p}(\textbf{q}_\text{p})}{3k_\text{B}},
\label{EQ12-Wave}
\end{equation}
where $\tilde{u}_\text{p}(\textbf{q}_\text{p})$ is the largest value among all eigenfunctions $\{\tilde{u}_a(\textbf{q})\}$ and values of $\textbf{q}$. We denote the wave vector at which this is found as $\textbf{q}_\text{p}$.

\begin{figure}
\centering
\includegraphics[clip,scale=0.62]{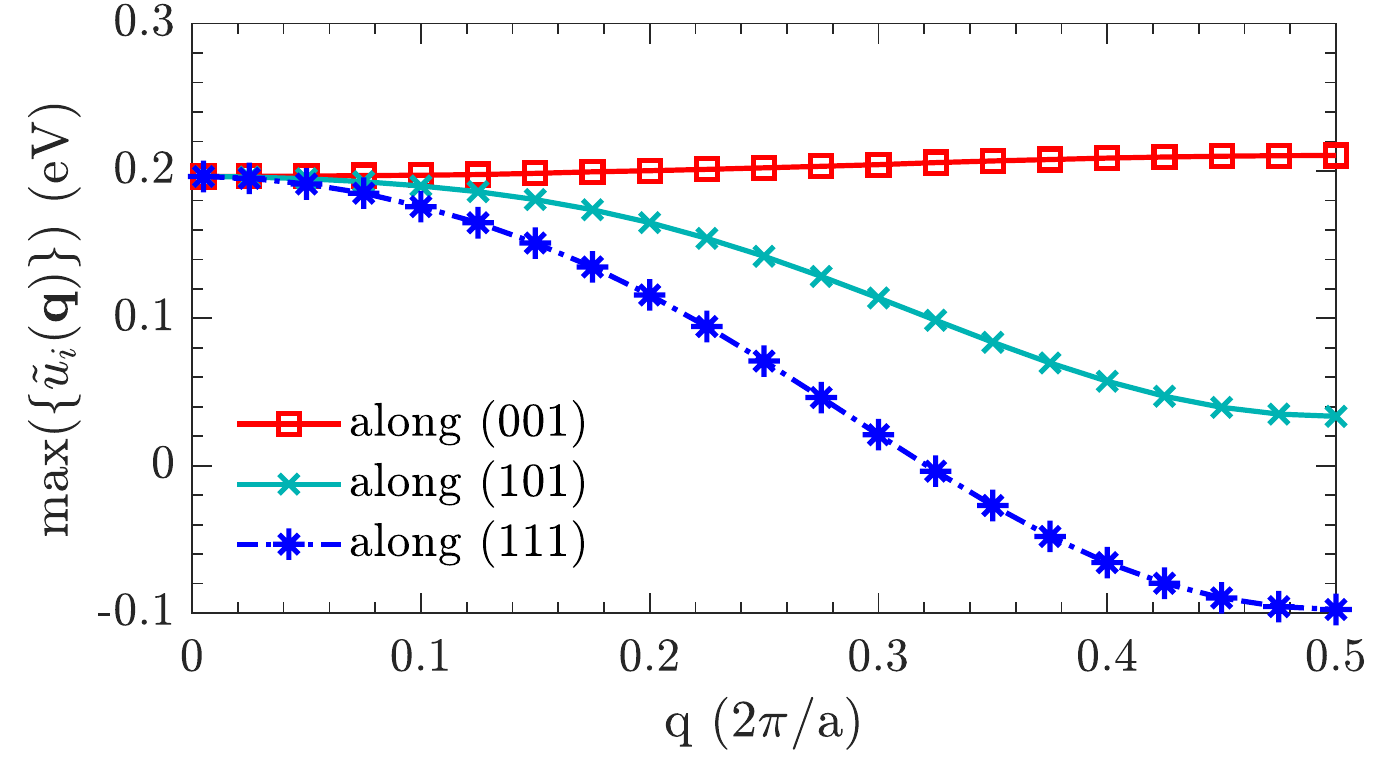}
\caption{Largest eigenvalue of $\tilde{\mathcal{S}}_{ss'}^{(2)}(\textbf{q})$ of cubic Mn$_3$Pt versus the wave vector $\textbf{q}$ for three characteristic directions in the reciprocal space and for a lattice parameter $a=3.95$\AA.}
\label{Mn3Pt_wave}
\end{figure}

If the temperature is below but close to $T_\text{max}$ the system develops some finite and small magnetic order described by $N_\text{at}$ local order parameters inside a unit cell $t$, $\{\delta\textbf{m}_{t1},\,\dots,\,\delta\textbf{m}_{tN_\text{at}}\}$, whose directions vary from one unit cell to another following the wave modulation of $\textbf{q}$.
The eigenvector components $\{V_{a,s}(\textbf{q})\}$ of $\tilde{\mathcal{S}}^{(2)}_{ss'}$, for a given eigenvalue $\tilde{u}_a(\textbf{q})$, contain the information of the relative orientations between the $N_\text{at}$ magnetic order parameters inside one unit cell $t$.
For example, if $\textbf{q}_\text{p}=\textbf{0}$ the PM state is unstable to the formation of ferromagnetic sub-lattices $s$, i.e. $\{\delta\textbf{m}_{ts}\}$ do not rotate from one unit cell to another. Note that $\{\delta\textbf{m}_{ts}\}$ form a ferromagnetic state only if all the components of the eigenvector are equal.
For illustrative purposes, in Fig.\ \ref{Mn3Pt_wave} we show some results obtained for Mn$_3$Pt in its cubic structure ($N_\text{at}=3$), which are described in more detail in section \ref{Mn3Pt}. An inspection of this figure reveals that two different magnetic phases compete in stabilization, corresponding to $\textbf{q}=\textbf{0}$ and $\textbf{q}=(\frac{1}{2},0,0)\frac{2\pi}{a}$, where $a$ is the lattice parameter. The components of the eigenvector are cosines perfectly matching a perfect triangular arrangement for $\textbf{q}=\textbf{0}$. Moreover, these components are $(+1,-1,0)$ for $\textbf{q}=(0,0,\frac{1}{2})\frac{2\pi}{a}$. Whilst in the first situation the local order parameters do not rotate from one unit cell to another, for $\textbf{q}=(0,0,\frac{1}{2})\frac{2\pi}{a}$ the wave vector is on the Brillouin zone edge and so $\{\delta\textbf{m}_{ts}\}$ completely reverse their orientations when changing from one unit cell to the next one along the $\hat{z}$ direction, and maintain the same orientation when transferred along the $\hat{x}$ and $\hat{y}$ directions, as shown in panels (a) and (b) of Fig.\ \ref{Mn3Pt1}, respectively. Longer periods occur when $\textbf{q}_\text{p}$ lies inside the Brillouin zone.

The calculation of $\{\textbf{h}^\text{int}_n\}$ and $\mathcal{G}_1$ for potential magnetic phases is guided by firstly studying the magnetic correlations in the PM state, described by $\tilde{\mathcal{S}}^{(2)}_{ss'}(\textbf{q})$~\cite{0305-4608-15-6-018,PhysRevLett.82.5369}. We inspect at which value $\textbf{q}_\text{p}$ the eigenvalues of $\tilde{\mathcal{S}}^{(2)}_{ss'}(\textbf{q})$ peak, from which we identify those magnetic phases to which the PM state is unstable when lowering the temperature. These magnetic states are described by the relative orientations between the local order parameters inside a unit cell, directly given by the eigenvector components, and by the modulation of $\textbf{q}_\text{p}$.
If Eq.\ (\ref{EQ1a-Wave}) and $\langle\Omega^\text{int}\rangle_0$ can be expressed in terms of $\mathcal{S}^{(2)}_{ij}$ (hence $\tilde{\mathcal{S}}^{(2)}_{ss'}(\textbf{q})$) completely, i.e.\ local moment direct pair correlations, only second-order magnetic phase transitions can be found.
First-order transitions are produced by higher order correlations, which have not been considered in this section ($f^{(a>2)}$ terms in Eq.\ (\ref{EQ1})). In section \ref{Multi} we explain their effect and how to obtain them.

\subsection{A hierarchy of local moment correlations}
\label{Multi}

The motion of the local moments is determined by the electronic structure and so their interactions stem directly from its complexity.
It is therefore unsurprising that in metallic magnetic materials such interactions cannot be only described in simple pairwise terms, but must also include interactions over groups of sites.
Many authors in the past have found theoretical evidence of the existence and importance of terms beyond a simple Heisenberg picture~\cite{PhysRevB.70.125115,PhysRevB.82.180404,PhysRevLett.107.017204,PhysRevLett.111.127204,PhysRevLett.116.217202}.
An indication of their presence follows directly from the calculation of effective pairwise interactions which adopt different values when evaluated from different magnetic states. For example, different results have been obtained in the ferromagnetic (FM) state compared to those in the PM state~\cite{0953-8984-19-32-326218,PhysRevB.72.104437} and some antiferromagnetic (AFM) phases~\cite{PhysRevB.77.174429}, as well as significant differences between PM and ferrimagnetic states~\cite{PhysRevB.93.184404}.
Multi-spin interactions have been demonstrated to be fundamental also in thin films, as predicted for the Mn monolayer on Cu(111)~\cite{PhysRevLett.86.1106} and for an up-up-down-down AFM state formed in monolayer Fe on bulk Rh(111)~\cite{PhysRevLett.120.207202,HoffmannArxiv}.
%

As explained, $\Omega_c(\{\hat{e}_n\})$, and therefore $\Omega^\text{int}(\{\hat{e}_n\})$, are likely to have a non-trivial dependence on the local moment orientations. The ensemble average $\langle\Omega^\text{int}\rangle_0$ can be, consequently, a very complicated function of the local moment magnetic order parameters $\{\textbf{m}_n\}$ too. We can express $\langle\Omega^\text{int}\rangle_0$ in the most general form as a sum of terms,
\begin{equation}
\langle\Omega^\text{int}\rangle_0=\Omega_0+f^{(2)}\left(\{\textbf{m}_n\}\right)+f^{(4)}\left(\{\textbf{m}_n\}\right)+\cdots,
\label{EQ1-Multi}
\end{equation}
where $\Omega_0$ is a constant and $f^{(a)}\left(\{\textbf{m}_n\}\right)$ is a $a$-th order function of $\{\textbf{m}_n\}$. These are the functions introduced in Eq.\ (\ref{EQ1}). Note that Eq.\ (\ref{EQ1-Multi}) remains unchanging under sign inversion of $\{\textbf{m}_n\}$ as these are axial vectors (they change sign under time reversal). If the magnetic state is isotropic, i.e.\ neglecting spin-orbit effects, the second and fourth order functions, for example, are expressed as
\begin{eqnarray}
\label{EQ2-Multia}
& & f^{(2)}\left(\{\textbf{m}_n\}\right)=-\frac{1}{2}\sum_{ij}\mathcal{S}^{(2)}_{ij}\textbf{m}_{i}\cdot\textbf{m}_{j} \\
& & f^{(4)}\left(\{\textbf{m}_n\}\right)=-\frac{1}{8}\sum_{ijkl}\mathcal{S}^{(4)}_{ijkl}\left(\textbf{m}_{i}\cdot\textbf{m}_{j}\right)\left(\textbf{m}_{k}\cdot\textbf{m}_{l}\right),
\label{EQ2-Multi}
\end{eqnarray}
where $\mathcal{S}^{(2)}_{ij}$ is the direct local moment pair correlation function introduced in Eq.\ (\ref{EQ1-Wave}) and the $\mathcal{S}^{(4)}_{ijkl}$ is the   direct quartet correlation function.  Eq.\ (\ref{EQ1-Multi}) shows $\mathcal{S}^{(4)}_{ijkl}$ to be defined formally as the fourth order derivative of $\langle\Omega^\text{int}\rangle_0$ with respect to the local magnetic order parameters.
We limit ourselves here to magnetic isotropic systems for simplicity, although our DFT-DLM codes are prepared to handle fully relativistic calculations and so anisotropy effects on $f^{(n)}\left(\{\textbf{m}_n\}\right)$ are also available~\cite{PhysRevB.95.155139}.

To illustrate the effect of $\mathcal{S}^{(4)}_{ijkl}$ and higher order correlations we study the dependence of the internal magnetic fields on $\{\textbf{m}_n\}$ given by Eq.\ (\ref{EQ14-Theo}) and so rewrite Eq.\ (\ref{EQ1a-Wave}) specifying the form of higher order terms,
\begin{equation}
\textbf{h}^\text{int}_i=-\sum_{ij}\mathcal{S}^{(2)}_{ij}\textbf{m}_j-\sum_{jkl,l>k}\mathcal{S}^{(4)}_{ijkl}\textbf{m}_j(\textbf{m}_k\cdot\textbf{m}_l)-\cdots.
\label{EQ3-Multi}
\end{equation}
Close to the PM phase the order at every magnetic site is very small ($\{\textbf{m}_n\}\rightarrow\{\textbf{0}\}$) and the local internal magnetic fields are, therefore, well described in terms of the $\mathcal{S}^{(2)}_{ij}$ which would be the exchange interactions, $J_{ij}$ for a simple pair-wise Heisenberg local moment model. However, as magnetic order increases, by reducing the temperature or applying an external magnetic field for example, $\{\textbf{m}_n\}$ increase in size and so higher order terms become important and must be considered in Eqs.\ (\ref{EQ1-Multi}) and (\ref{EQ3-Multi}). In other words, the presence of beyond pairwise Heisenberg-like interactions among the local moments, $\{\hat{e}_n\}$, has the effect of changing the magnetic behavior as the ordering develops at each site, which is reflected from a non-linear nature of $\textbf{h}_n^\text{int}$ in Eq.\ (\ref{EQ3-Multi}).
The meaning of this is that thermal fluctuations of $\{\hat{e}_n\}$ at different states of magnetic order can induce alterations on the electronic structure. As it spin-polarizes, the electronic structure transforms and in turn affects the magnetic interactions between the local moments, which is captured by the importance of higher order direct local moment correlations, $\mathcal{S}^{(4)}_{ijkl}$ etc. in the free energy.
This itinerant electron effect can underly the origin of first-order magnetic phase transitions~\cite{PhysRevLett.118.197202}.

\textit{Calculation of higher order correlations:} We can obtain $\mathcal{S}^{(2)}_{i=\{ts\}j=\{t's'\}}$ from $\tilde{\mathcal{S}}^{(2)}_{ss'}(\textbf{q})$ and for a set of $\textbf{q}$'s using the lattice transform in Eq.\ (\ref{EQ7-Wave3}).
The method to calculate $\mathcal{S}^{(4)}_{ijkl}$ and higher order terms, moreover, consists in producing many \textit{ab initio} local field data $\{\textbf{h}^\text{int}_n\}$ for different selected values of $\{\textbf{m}_n\}$. We propose a function for the non-linear terms in Eq.\ (\ref{EQ3-Multi}) exploiting the symmetries of the magnetic states under study.
This results in a reduced number of constants that compactly contain all significant and relevant higher order terms.
Then a least squares fitting is performed into the proposed function to test it and to extract the constants. A crucial point is to thoroughly scan the magnetic phases of interest and verify that constants are not over fitted, which in turn determines the function proposed~\cite{PhysRevB.89.054427,PhysRevLett.115.207201,PhysRevLett.118.197202,PhysRevB.95.184438}.  

To illustrate this we present here the case of a simple ferromagnet, i.e.\ $\{\textbf{m}_n=\textbf{m}_\text{FM}\}$ with $\tilde{\mathcal{S}}^{(2)}_{s=1s'=1}(\textbf{q})\equiv\tilde{\mathcal{S}}^{(2)}(\textbf{q})$ being of dimension $N_\text{at}=1$. Eq.\ (\ref{EQ3-Multi}) then becomes
\begin{equation}
\textbf{h}^\text{int}_i+\mathcal{S}^{(2)}_\text{FM}\textbf{m}_\text{FM}=-\mathcal{S}^{(4)}_\text{FM}m_\text{FM}^2\textbf{m}_\text{FM}-\cdots
\label{EQ4-Multi}
\end{equation}
where
\begin{equation}
  \mathcal{S}^{(2)}_\text{FM}=\tilde{\mathcal{S}}^{(2)}(\textbf{0})=\sum_{j}\mathcal{S}^{(2)}_{ij}, \,\,\,\,\,\, \mathcal{S}^{(4)}_\text{FM}=\sum_{jkl,l>k}\mathcal{S}^{(4)}_{ijkl} 
\label{EQ5-Multi}
\end{equation}
The function to be fitted is the right hand side of Eq.\ (\ref{EQ4-Multi}). The calculation of $\textbf{h}^\text{int}_n$ as a function of $\textbf{m}_\text{FM}$ is therefore sufficient to determine the fitted function and so extract $\mathcal{S}^{(4)}_\text{FM}$, and higher order terms.
The generalization of this procedure to more complicated magnetic states is used for the non-collinear AFM and other magnetic phases in Mn$_3$A class of materials. In section \ref{Mn3A} we will show in which systems high order terms are important.

\section{Free energy minimization: Magnetic phase diagrams and caloric effects}
\label{minim}

Our DFT-DLM computational codes directly provide the first derivative of $\langle\Omega^\text{int}\rangle_0$, i.e.\ the local fields $\{\textbf{h}^\text{int}_n\}$ and the direct pair local moment correlation functions $\tilde{\mathcal{S}}^{(2)}_{ss'}(\textbf{q})$, respectively (appendix \ref{S2APP}).
By repeating the calculation of these quantities at different lattice structure values, the magnetic interactions can be obtained as functions of the lattice deformation in order to account for a magnetoelastic effect, i.e.\ $\tilde{\mathcal{S}}^{(2)}_{ss'}(\textbf{q},\bm{\varepsilon}_{\alpha\beta})$ (or $\mathcal{S}^{(2)}_{ij}(\bm{\varepsilon}_{\alpha\beta})$), $\mathcal{S}^{(4)}_{ijkl}(\bm{\varepsilon}_{\alpha\beta})$, $\dots$, where $\bm{\varepsilon}_{\alpha\beta}$ is the strain tensor~\cite{LandauElasticity}.
Importantly, we obtain self-consistent KKR-DFT potentials at different values of $\bm{\varepsilon}_{\alpha\beta}$. The dependence of the local moment sizes on the lattice deformation, $\{\mu_n(\bm{\varepsilon}_{\alpha\beta})\}$, and its effect on the magnetic interactions, is included in consequence.

To obtain our central expression given in Eq.\ (\ref{EQ1}), we add to Eq.\ \ref{EQ11-Theo} the terms $\frac{1}{2}C_{\alpha\beta\gamma\kappa}\bm{\varepsilon}_{\alpha\beta}\bm{\varepsilon}_{\gamma\kappa}$ and $\sigma_{\alpha\beta}\bm{\varepsilon}_{\alpha\beta}$, which are the simplest elastic term describing the dependence of the total energy on $\bm{\varepsilon}_{\alpha\beta}$ and the effect of stress application $\sigma_{\alpha\beta}$, respectively. $C_{\alpha\beta\gamma\kappa}$ can be expressed in terms of the inverse of the compressibility or the shear and Young's modulus, for example, and can be calculated from DFT calculations or directly taken from experiment.

Once $\tilde{\mathcal{S}}^{(2)}_{ss'}(\textbf{q},\bm{\varepsilon}_{\alpha\beta})$, $\mathcal{S}^{(4)}_{ijkl}(\bm{\varepsilon}_{\alpha\beta})$ and higher order terms are obtained, $\mathcal{G}_1$ can be computed from direct application of Eq.\ (\ref{EQ1}). At this point a trivial numerical minimization of $\mathcal{G}_1$ is performed to obtain the equilibrium properties.
We construct magnetic phase diagrams by calculating and comparing $\mathcal{G}_1$ of magnetic structures of interest. The magnetic ordering $\{\textbf{m}_n\}$ that globally minimizes $\mathcal{G}_1$ is considered as the most stable phase at every point in the diagram, defined at different values of the temperature and other parameters, such as the strength of an external magnetic field and lattice structure values.

\subsection{Electronic and magnetoelastic origin of fourth order coupling}
\label{MSE}

We focus on the situation in which such a magnetoelastic effect has a significant impact on the leading direct pair correlations only and assume a linear dependence due to small deformations,
\begin{equation}
\mathcal{S}^{(2)}_{ij}(\bm{\varepsilon}_{\alpha\beta})\approx\mathcal{S}^{(2)}_{0,ij}+\alpha_{ij}\bm{\varepsilon}_{\alpha\beta},
\label{EQ1-MSE}
\end{equation}
where $\mathcal{S}^{(2)}_{0,ij}$ are obtained for some reference value of the relaxed lattice structure in the PM state, and $\alpha_{ij}$ are constants describing the magnetoelastic coupling.
To illustrate the effect of $\alpha_{ij}$ we proceed to minimize Eq.\ (\ref{EQ1}) with respect to $\bm{\varepsilon}_{\alpha\beta}$. For simplicity we restrict ourselves to study a mechanical system under application of a hydrostatic pressure $p$, and so set $\frac{1}{2}C_{\alpha\beta\gamma\kappa}\bm{\varepsilon}_{\alpha\beta}\bm{\varepsilon}_{\gamma\kappa}=\frac{1}{2}V_0\gamma\omega^2$ and $\sigma_{\alpha\beta}\bm{\varepsilon}_{\alpha\beta}=pV_0\omega$, where $V_0$ is the volume of the unit cell, $\gamma$ is the inverse of the compressibility, and $\omega=(V-V_0)/V_0$ is the relative volume change. From $\partial\mathcal{G}_1/\partial\omega=0$ we obtain
\begin{equation}
\omega=\frac{1}{\gamma}\left(\frac{1}{2V_0}\sum_{ij}\alpha_{ij}\textbf{m}_i\cdot\textbf{m}_j-p\right),
\label{EQ3-MSE}
\end{equation}
which substituted into Eq.\ (\ref{EQ1}) gives
\begin{widetext}
\begin{equation}
 \mathcal{G}_1=-T\sum_n S_n(\textbf{m}_n) + \Bigg[\Omega_0-\frac{1}{2}\sum_{ij}\left(\mathcal{S}^{(2)}_{0,ij}-\frac{p\alpha_{ij}}{\gamma}\right)\textbf{m}_i\cdot\textbf{m}_j
 -\frac{1}{8}\sum_{ijkl}\mathcal{S}^{(4)}_{ijkl}(\textbf{m}_i\cdot\textbf{m}_j)(\textbf{m}_k\cdot\textbf{m}_l)-h.o.-\frac{\left(\sum_{ij}\alpha_{ij}\textbf{m}_i\cdot\textbf{m}_j\right)^2}{8V_0\gamma}-\frac{p^2V_0}{\gamma}\Bigg].
\label{EQ4-MSE}
\end{equation}
\end{widetext}
where $h.o.$ stands for higher order local moment correlations. Hence, a magnetoelastic coupling $\alpha_{ij}$ in general produces a fourth order contribution to the free energy, of biquadratic form, to add to the fourth order term of electronic origin $\mathcal{S}^{(4)}_{ijkl}$.
In general both contributions can be present and so be the driving factor of a first-order PM-ordered and/or any ordered-to-ordered magnetic phase transition. For example, we have shown that $\mathcal{S}^{(4)}_{ijkl}$ play a crucial role on phase stabilization in the heavy rare earth elements~\cite{PhysRevLett.118.197202} and some gadolinium intermetallics~\cite{PhysRevLett.115.207201}, and that both $\mathcal{S}^{(4)}_{ijkl}$ and $\alpha_{ij}$ are essential to explain the origin of first-order PM-to-triangular AFM phase transitions in Mn-based antiperovskites~\cite{PhysRevB.95.184438,PhysRevX.8.041035}. In section \ref{Mn3Pt} we will show how these two sources contribute to the pressure-temperature magnetic phase diagram of Mn$_3$Pt and that $\mathcal{S}^{(4)}_{ijkl}$ and higher order correlations are in general present in Mn$_3$A.

\subsection{Caloric effects}
\label{CE}

Caloric responsive materials show a substantial change of their thermodynamic state when an external field of some sort is applied and/or removed~\cite{Tishin,PlanesSaxena2014,SANDEMAN2012566}.
Depending on the external stimulus triggering the change, the effect is called magnetocaloric (MCE)~\cite{PhysRevLett.78.4494}, barocaloric (BCE)~\cite{Matsunami1,Lloveras1}, electrocaloric (ECE)~\cite{Neese2008,ADMA:ADMA201203823}, elastocaloric (eCE)~\cite{PhysRevLett.100.125901}, and toroidocaloric (TCE)~\cite{0953-8984-20-43-434203,PhysRevB.85.144429,PhysRevB.84.094421,PhysRevB.88.024414}, for magnetic field, hydrostatic pressure, electric field, uniaxial/biaxial stress, and toroidic field, respectively. Note that the BCE and eCE effects are particular cases driven by mechanical stresses, often referred to as mechanocaloric effects.
Magnetic refrigeration based on the exploitation of one or multiple caloric effects has become a widely investigated technology and promises to be an environment friendly and more efficient alternative to gas-compression based devices at room temperature~\cite{Tishin,PlanesSaxena2014}.
The magnetic refrigeration community, however, has little guidance from theoretical research and its scientific advances are often heuristic in nature. The DLM theory presented here is designed to naturally evaluate from first-principles entropy changes at finite temperatures. Crucially, the capability of the approach to characterize the order of the transitions as well as to predict and locate tricritical points make the theory suitable to identify optimal cooling cycles from constructed phase diagrams~\cite{PhysRevB.87.060404,PhysRevB.89.054427,PhysRevB.95.184438}.

In general, a caloric effect is quantified by the isothermal entropy change, $\Delta S_{iso}$, and the adiabatic temperature change, $\Delta T_{ad}$, induced in the thermodynamic conjugate of the external field applied and/or removed.
Our DLM theory can directly provide the total entropy as a function of the state of magnetic order, as well as the dependence on temperature, magnetic field and crystal structure. It naturally predicts the entropy changes due to the orientational disorder of the local moments, $\Delta S_\text{mag}=\Delta(\sum_n S_n)$, from Eq.\ (\ref{EQ10-Theo})~\cite{0953-8984-26-27-274210}. We can also estimate the entropy change from alterations of the electronic density of states, supporting and generating the local moments, by using the Sommerfeld expansion~\cite{AshcroftMermin}
\begin{equation}
S_{elec}=\frac{\pi^2}{3}k_\text{B}^2Tn(\{\textbf{m}_n\},E_\text{F}),
\label{EQ1-Cal}
\end{equation}
where the electronic density $n(\{\textbf{m}_n\},E_\text{F})$ is given at the Fermi energy and depends on the state of magnetic order. We label this contribution electronic entropy, and due to the nature of the time scale separation between slow varying local moment orientations and fast underlying electronic motions, it is formally captured within the internal energy
\begin{equation}
\langle\Omega^\text{int}\rangle_0=\bar{E}-TS_{elec},
\label{EQ2-Cal}
\end{equation}
where $\bar{E}$ and $S_{elec}$ are the DFT-based energy and electronic entropy averaged over local moment configurations~\cite{PhysRev.137.A1441,PhysRevB.89.054427}.
Of course, the entropy change $S_\text{mag}+S_{elec}$ is entirely electronic in origin. While $S_\text{mag}$ captures the contribution from the long-lived local moments emerging from the interacting electrons, $S_{elec}$ captures that from the remaining faster electronic modes.

Following the Born-Oppenheimer approximation it would be reasonable to consider the lattice vibrations to fluctuate on the slowest time scale $\tau_\text{vib}\gg\tau_\text{mag}\gg\tau_\text{elec}$. Under these circumstances it should be possible to use the same ideas developed for the magnetic fluctuations and expand the theory to incorporate the effect of the vibrational fluctuations. Such a theory could address the entire magneto-phonon coupling at finite temperatures.
Here, however, the atomic positions are fixed and this effect is not taken into account. We follow an alternative and simpler approximation to incorporate the effect of the lattice vibrations via the implementation of a straightforward Debye model, defining the vibrational entropy as~\cite{Tishin,Gopal}  
\begin{equation}
S_{vib} =  k_\text{B} \Biggl[-3\ln\left(1-e^{-\frac{\theta_\text{D}}{T}}\right)+12\left(\frac{T}{\theta_\text{D}}\right)^{3}\int_{0}^{\frac{\theta_\text{D}}{T}}\frac{x^{3}}{e^{x}-1}dx\Biggr],
\label{EQ3-Cal}
\end{equation}
where $\theta_\text{D}$ is the Debye temperature, which can be obtained from experiment or other first principles sources~\cite{CHEN2001947}. The presence of this term acts purely as a thermal bath or reservoir to exchange entropy with the electronic and magnetic degrees of freedom.
Hence, in our approach the total entropy is directly given as
\begin{equation}
S_{tot}=S_\text{mag}+S_{elec}+S_{vib}.
\label{EQ3a-Cal}
\end{equation}
This is the central equation in our method for the calculation of caloric effects. For example, under the presence of an external magnetic field that varies as $\textbf{H}=\textbf{H}_0\rightarrow\textbf{H}_1$, we compute the isothermal entropy change from
\begin{equation}
\Delta S_{iso}(T,\textbf{H}_0\rightarrow\textbf{H}_1)=S_{tot}(T,\textbf{H}_1)-S_{tot}(T,\textbf{H}_0).
\label{EQ4-Cal}
\end{equation}
Similarly, we obtain the adiabatic temperature change by solving the equation
\begin{equation}
S_{tot}(T,\textbf{H}_0)=S_{tot}(T+\Delta T_{ad},\textbf{H}_1).
\label{EQ5-Cal}
\end{equation}
In addition, mechanocaloric effects can be estimated by the calculation of thermal responses caused by the change of the lattice structure, triggered either by application of a hydrostatic pressure or a mechanical stress, 
\begin{eqnarray}
& \Delta S_{iso}(T,\bm{\varepsilon}_{0\alpha\beta}\rightarrow\bm{\varepsilon}_{\alpha\beta})=S_{tot}(T,\bm{\varepsilon}_{\alpha\beta})-S_{tot}(T,\bm{\varepsilon}_{0\alpha\beta}),\,\,\,\,\,\,\,\,\,\, \\
& S_{tot}(T,\bm{\varepsilon}_{0\alpha\beta})=S_{tot}(T+\Delta T_{ad},\bm{\varepsilon}_{\alpha\beta}),
\label{EQ6-Cal}
\end{eqnarray}
where $\bm{\varepsilon}_{0\alpha\beta}$ and $\bm{\varepsilon}_{\alpha\beta}$ stand for the strain tensor before and after the stress application, respectively. From these equations the theory is able to provide field-tuned and multicaloric effects involving magneto- and mechano- caloric responses~\cite{PhysRevB.89.054427,PhysRevB.95.184438}.
Conventional/inverse caloric effects, in which cooling is based on adiabatic demagnetization/magnetization when an external magnetic field is removed/applied can be modeled by our approach.
Some examples in which inverse caloric effects are present are the off-stoichiometry FeRh system~\cite{NIKITIN1992234,PhysRevB.89.214105,PhysRevB.89.054427}, the FIM-AFM transition in doped Mn$_2$Sb compounds~\cite{doi:10.1063/1.117218,doi:10.1063/1.4821197}, the non-collinear magnetism in Mn$_5$Si$_3$~\cite{TEGUS2002174}, and the metamagnetic magnetoelastic transition in CoMnSi~\cite{PhysRevB.74.224436,PhysRevB.87.064410}.

\section{Temperature-dependence of frustrated magnetism in M\lowercase{n}$_3$A}
\label{Mn3A}

The Mn$_3$A family of itinerant magnetic materials shows a very rich range of magnetic phases~\cite{0953-8984-25-20-206006} which are currently intensively studied for their AFM~\cite{PhysRevLett.119.187204,PhysRevB.95.075128,PhysRevLett.112.017205} and ferrimagnetic (FIM)~\cite{PhysRevB.83.020405,YOU201740,doi:10.1063/1.2722206,doi:10.1063/1.4970691,GUTIERREZPEREZ201820,GUTIERREZPEREZ201714} properties, and potential for spintronic applications. The element A can be one among many (Pt, Ir, Rh, Sn, Ga, Ge). Three distinct lattice structures crystallize depending on the element A, namely cubic for A=Pt, Ir, Rh being a transition metal, and hexagonal for A=Sn, Ga, Ge, and tetragonal for A=Ga, Ge. Despite the different electronic structures produced by the different elements A, a common trait is that the Mn atoms are positioned such that triangular and pyramid atomic connections are formed in all systems (see Figs.\ \ref{Mn3Pt1}, \ref{FigHexagonal}, and \ref{Mn3Ga1_tog}(a)). Evidently, this situation is likely to generate geometrically frustrated AFM interactions. For example, non-collinear AFM triangular states arise from the formation of Kagome-type lattice planes in both the cubic and hexagonal lattices~\cite{PhysRevB.95.075128}, whilst the tetragonal systems show collinear FIM~\cite{KREN19701653,PhysRevB.77.054406,PhysRevB.93.184404}.
The material that perhaps exhibits the most intriguing frustrated magnetism is cubic Mn$_3$Pt, in which the high temperature collinear AFM phase has been observed experimentally to have an apparent zero net magnetization at some Mn sites~\cite{PhysRev.171.574,doi:10.1143/JPSJ.56.4532,doi:10.1063/1.2163485}.
It is a challenging test of our DLM theory to provide an explanation for this apparent oddity as well as the disparate magnetic properties of the other Mn$_3$A systems.

We perform comprehensive DFT-DLM calculations to study the effect of thermal fluctuations on all observed magnetic phases in Mn$_3$A, which contain complex multi-atom sub-lattices per unit cell, and how the spin-polarized electronic structure consequently reacts to the change of magnetic order and its link to frustration.
As the temperature is raised, fluctuations of the local moment orientations increase. If the magnetic interactions permit, the theory can model fully frustrated magnetic sites in which there is no net spin polarization after averaging over all possible directions, although the local moment size is yet stabilized by a local exchange splitting. In our theory this situation is described by the magnetic local order parameter of the corresponding sub-lattice being zero ($m_n=0$). This is a key aspect to describe the magnetism of Mn$_3$A and something inaccessible by standard zero-temperature DFT calculations constrained to describe fully ordered magnetic states ($\{m_n=1\}$).

For each material we follow the methodology explained in section \ref{Theory}: We calculate the direct correlation function in the reciprocal space $\tilde{\mathcal{S}}^{(2)}_{ss'}(\textbf{q})$, describing the magnetic correlations in the PM state, and identify the potential magnetic phases that can stabilize. This calculation is then used together with the more detailed study based on the local magnetic fields $\{\textbf{h}^\text{int}_n\}$ sustaining the local moments at different magnetic orderings.
Section \ref{Mn3Pt} focuses on the situation in which A is a transition metal.
In section \ref{NTMSect} we study the magnetism when A=Sn, Ge, and Ga. Results for the triangular AFM state in the hexagonal structure, and the FIM state in the tetragonal structure, are shown in sections \ref{hexa} and \ref{tetra}.
We use lattice parameter values directly taken from experiment.

\subsection{First- and second-order magnetic phase transitions in cubic Mn$_3$A (A=Pt, Ir, Rh)}
\label{Mn3Pt}

\subsubsection{Experimental properties}

\begin{figure}[t]
\centering
\includegraphics[clip,scale=0.28]{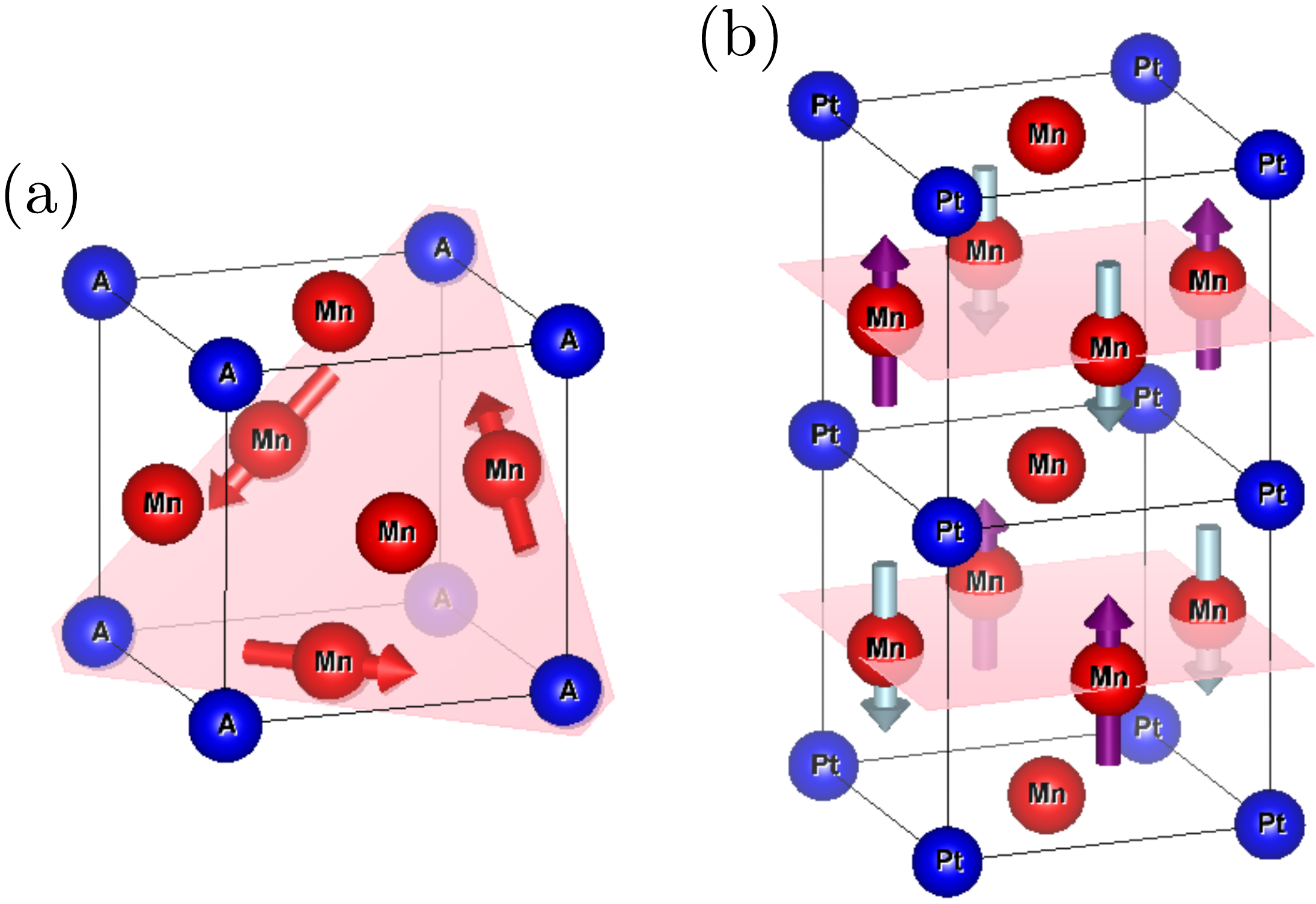}
\caption{Magnetic unit cells of the (a) triangular AFM and (b) collinear AFM magnetic states in cubic Mn$_3$A. Arrows are used to indicate the net magnetic moment after performing the average over the local moment orientations, i.e.\ the local order parameters $\{\textbf{m}_n\}$. Note that for the collinear AFM phase the magnetic sites outside the pink layers have $\textbf{m}_n=\textbf{0}$.}
\label{Mn3Pt1}
\end{figure}

When A is a transition metal Mn$_3$A crystallizes into the Cu$_3$Au-type cubic structure, where the Mn atoms are located at the face centers and A atoms occupy the corner positions. Whilst Mn$_3$Ir and Mn$_3$Rh show a second-order transition from the PM state to a triangular AFM state (Fig.\ \ref{Mn3Pt1}(a))~\cite{doi:10.1063/1.371298,KREN1966331}, Mn$_3$Pt shows a second-order collinear AFM-PM transition when cooling through $T_N=475$K (Fig.\ \ref{Mn3Pt1}(b))~\cite{PhysRev.171.574,doi:10.1143/JPSJ.56.4532,doi:10.1063/1.371298,doi:10.1063/1.2163485}. The corresponding magnetic unit cell of the collinear AFM phase is twice (and so tetragonal) the crystallographic unit cell. Sites with non-zero net moment form AFM planes, indicated by pink layers in the figure, stacked perpendicular to the $c$ axis and modulated with a wave vector $\textbf{q}=(0, 0, 0.5)\frac{2\pi}{a}$, where $a$ is the lattice parameter. The local order parameters describing this state, therefore, completely invert their orientations from one layer to the adjacent ones. The Mn atoms with vanishing net local magnetic order sit in layers staggered between the AFM planes.

Mn$_3$Ir and Mn$_3$Rh do not exhibit another transition and so their triangular AFM state remains stable at lower temperatures. However, reducing the temperature further down to $T_{tr}=365$K~\cite{doi:10.1143/JPSJ.56.4532,doi:10.1063/1.2163485} in Mn$_3$Pt triggers an additional first-order transition from the collinear to the triangular AFM state.
An important observation is that the transition temperatures of Mn$_3$Pt change significantly under pressure application with reported values as $\text{d}T_{N}/\text{d}p=-70$K/GPa and $\text{d}T_{tr}/\text{d}p=140$K/GPa~\cite{doi:10.1143/JPSJ.56.4532}. A hydrostatic pressure of $p_c\approx 0.3$GPa is in consequence enough to completely destroy the collinear AFM order and a unique PM-triangular AFM phase transition is found for $p>p_c$, hence suggesting the presence of a considerable magnetovolume coupling. This is in line with the phenomenological magnetic phase diagram for pairwise interactions constructed by Shirai \textit{et al}.~\cite{SHIRAI1990157} to investigate the itinerant magnetism of Mn$_3$Pt. In this work the authors suggested that the change of magnetic interactions induced by applied pressures, i.e.\ a magnetovolume coupling, can produce the magnetic phase transitions observed in Mn$_3$Pt.
We will show that indeed a magnetovolume coupling is the primary origin of this transition.

\subsubsection{High temperature regime and magnetovolume coupling}
\label{highCub}

We firstly inspect the direct correlation function in the PM limit to see if the triangular AFM state for Mn$_3$Rh and Mn$_3$Ir, and the collinear AFM state for Mn$_3$Pt, are the potential stable magnetic phases. Since the crystallographic unit cell contains three Mn atoms, $\tilde{\mathcal{S}}_{ss'}^{(2)}(\textbf{q})$ is a 3$\times$3 matrix with three eigenfunctions $\{\tilde{u}_i(\textbf{q}), i=1,2,3\}$.
We carried out DFT-DLM calculations for the three magnetic materials at their respective experimental lattice parameters in the PM state. Local magnetic moments with sizes decreasing as $\mu_\text{Mn(Pt)}>\mu_\text{Mn(Rh)}>\mu_\text{Mn(Ir)}$ established at each Mn site. In table \ref{TabCub} we show these values and the experimental lattice parameters used in the calculations.

\begin{table}
\begin{center}
\begin{tabular}{ c | c c c c }
 \hline\hline
 & $a_\text{exp}$ (\AA) & $\mu_\text{Mn}$ ($\mu_\text{B}$) & $T_N^\text{theo}$ (K) & $T_N^\text{exp}$ (K)  \\
 \hline
Mn$_3$Pt  ~\cite{PhysRev.171.574}      & 3.87 & 3.47 & $\approx$790  & 475 \\
Mn$_3$Ir  ~\cite{doi:10.1063/1.371298} & 3.82 & 2.83 & 1300 & 960 \\
Mn$_3$Rh  ~\cite{KREN1966331}          & 3.81 & 3.41 & 1400 & 850 \\
 \hline\hline
\end{tabular}
\caption{Application of the theory to cubic Mn$_3$A (A=Pt, Ir, Rh). The table shows the lattice parameters used for each material, taken directly from experiment, and theory results for local moment sizes, and N\'eel transition temperatures, $T_N^\text{theo}$, which are compared with experimental values, $T_N^\text{exp}$.}
\label{TabCub}
\end{center}
\end{table}

\begin{figure}[t]
\centering
\includegraphics[clip,scale=0.62]{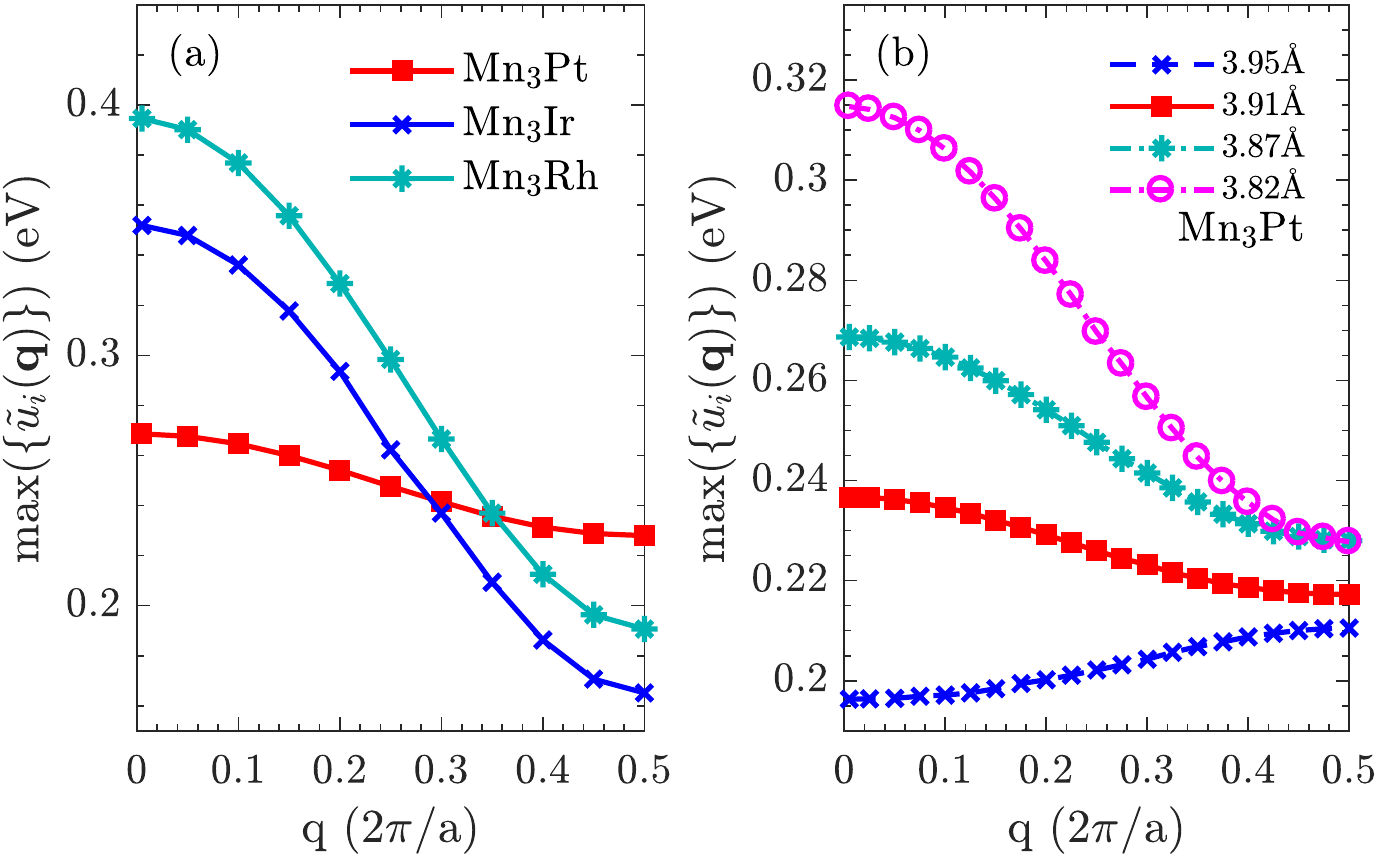}
\caption{Largest eigenvalue of the direct correlation function along the (001) direction in the reciprocal space for (a) Mn$_3$Pt (red squares), Mn$_3$Ir (blue crosses), and Mn$_3$Rh (turquoise stars) at experimental lattice parameters (see table \ref{TabCub}), and (b)for Mn$_3$Pt for a range of lattice parameters.}
\label{Mn3PtTOG}
\end{figure}

Fig.\ \ref{Mn3PtTOG}(a) shows the largest eigenvalue of $\tilde{\mathcal{S}}_{ss'}^{(2)}(\textbf{q})$ along the direction (001) in the reciprocal space. We explored the $\textbf{q}$-dependence of $\tilde{\mathcal{S}}_{ss'}^{(2)}(\textbf{q})$ and verified that there are no other potential magnetic phases, as illustrated for Mn$_3$Pt with $a=3.95$\AA\, in Fig.\ \ref{Mn3Pt_wave}. Fig.\ \ref{Mn3PtTOG}(a) shows that there are two competing $\textbf{q}$-points corresponding to $\textbf{q}=\textbf{0}$ and $\textbf{q}=(0, 0, 0.5)\frac{2\pi}{a}$.
Pleasingly, we have found that the eigenvector components of $\tilde{\mathcal{S}}_{ss'}^{(2)}(\textbf{q})$ at $\textbf{q}=\textbf{0}$ and $\textbf{q}=(0, 0, 0.5)\frac{2\pi}{a}$ are in direct agreement with the magnetic order found in experiment. For $\textbf{q}=(0, 0, 0.5)\frac{2\pi}{a}$ they adopt the shape of (++0), i.e.\ the collinear AFM state that stabilizes in Mn$_3$Pt (Fig.\ \ref{Mn3Pt1}(b)), and for $\textbf{q}=\textbf{0}$ the components are cosines which describe the triangular arrangement that stabilizes in the three materials (Fig.\ \ref{Mn3Pt1}(a)), i.e.\ $(1,-\frac{1}{2},-\frac{1}{2})$. Note that the first two components refer to Mn atoms positioned within the pink layers in Fig.\ \ref{Mn3Pt1}(b), i.e.\ at (0.5 0 0.5) and (0 0.5 0.5), while the third refers to the site with no net local magnetic moment orientational order, at (0.5 0.5 0), in units of lattice parameters.
Remarkably, results for Mn$_3$Pt are in sharp contrast to Mn$_3$Ir and Mn$_3$Rh (Fig.\ \ref{Mn3PtTOG}(a)). Whilst the peak at $\textbf{q}=(0, 0, 0.5)\frac{2\pi}{a}$, corresponding to the collinear AFM state, is strongly suppressed for Mn$_3$Ir and Mn$_3$Rh and the triangular structure is therefore the stable state as found in experiment, for Mn$_3$Pt both $\textbf{q}=\textbf{0}$ and $\textbf{q}=(0, 0, 0.5)\frac{2\pi}{a}$ have similar eigenvalues and so similar stability.
In table \ref{TabCub} we show the second-order transition temperatures obtained from the largest eigenvalues of $\tilde{\mathcal{S}}_{ss'}^{(2)}(\textbf{q})$, and their comparison with experiment.

Since both triangular AFM at $\textbf{q}=\textbf{0}$ and collinear AFM at $\textbf{q}=(0, 0, 0.5)\frac{2\pi}{a}$ are close in energy in Mn$_3$Pt, further analysis focused on the magnetovolume coupling and the effect of higher order local moment correlations is required for this material (see sections \ref{Multi} and \ref{minim}).
First, we investigate the magnetovolume effect by repeating calculations at different volumes, self-consistently including the effect of local moment magnitude change on the magnetic correlations. We obtained a linear dependence as $\mu_\text{Mn}=(3.47+3.3\frac{\Delta V}{V_0})\mu_\text{B}$ for Mn$_3$Pt, where $V_0$ and $\Delta V$ are the volume of the unit cell and its relative change.

Fig.\ \ref{Mn3PtTOG}(b) shows that the competition between the collinear and triangular AFM states can be strongly controlled by changing the volume, i.e.\ the lattice parameter $a$, indicating the presence of a large magnetovolume effect.
For increasing values of $a$, Mn$_3$Pt's paramagnetic correlations increasingly favor the collinear AFM state.
In fact, for $a=3.95$\AA\, the maximum eigenvalue of $\tilde{S}_{ss'}^{(2)}(\textbf{q})$ peaks at $\textbf{q}=(0, 0, 0.5)\frac{2\pi}{a}$ and the PM state is, therefore, unstable to the formation of the collinear AFM phase.
The absence of this state for A=Ir and Rh evidently is caused by the lattice contraction from the presence of these transition metals.
Note that the experimental lattice parameters of these two materials are significantly contracted compared to Mn$_3$Pt.
The paramagnetic correlations in Mn$_3$Ir and Mn$_3$Rh thus strongly favor the triangular AFM state.
This is further confirmed from the results shown in section \ref{MPD}, where we calculate the contribution from higher order correlations and complete our finite temperature study of Mn$_3$Pt. After minimizing the free energy we obtain a critical value of the lattice parameter $a_c$ such that structures with $a<a_c$ show a single PM-triangular AFM second-order phase transition for Mn$_3$Pt. We find that the collinear AFM state is not stable in Mn$_3$Ir and Mn$_3$Rh even for expansions around $a_c$.

\subsubsection{Lower temperature regime and the magnetic phase diagram of Mn$_3$Pt}
\label{MPD}

\begin{figure}[t]
\centering
\includegraphics[clip,scale=0.63]{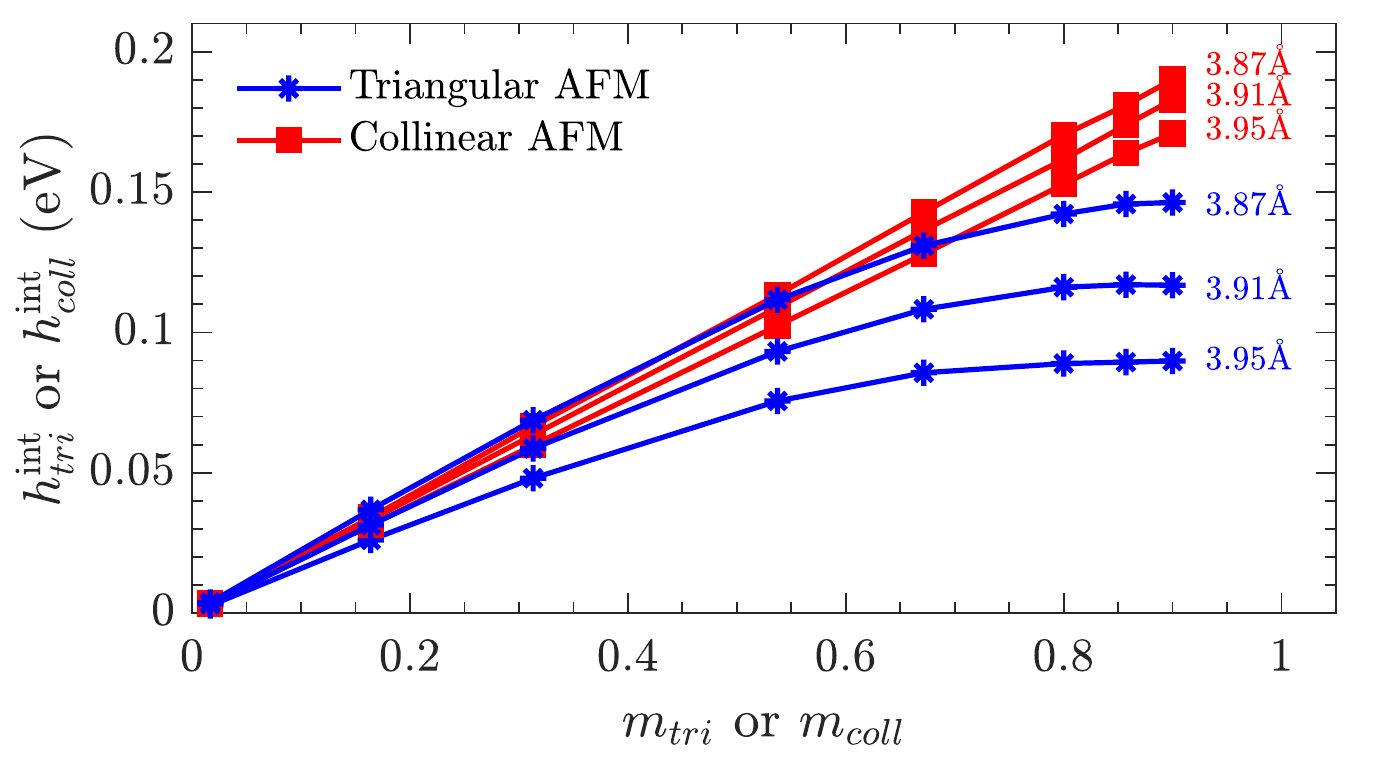}
\caption{Absolute value of the internal fields, $h^\text{int}_{tri}$ and $h^\text{int}_{coll}$, for the triangular (red squares) and collinear (blue stars) AFM states as functions of magnetic order and for a range of lattice parameters.}
\label{Mn3Pt3}
\end{figure}

We now use our DFT-based DLM theory to calculate the internal magnetic fields $\{\textbf{h}^\text{int}_n\}$ as functions of the local order parameter, extract the higher order local moment correlation functions and produce the free energy. Here the triangular AFM phase is formed by local order parameters forming 120 degrees, and so a single size $m_{tri}$ associated with each of the three Mn sites describes this magnetic state. Similarly, by symmetry the magnitudes among $\{\textbf{m}_n\}$ at sites with non-zero net magnetic moment are the same for the collinear state, that we label as $m_{coll}$. They have opposite directions as shown in Fig.\ \ref{Mn3Pt1}(b). We define $h^\text{int}_{tri}$ and $h^\text{int}_{coll}$ as the absolute values of the effective fields sustaining the local moments for the triangular and collinear AFM states, respectively. In Fig.\ \ref{Mn3Pt3} we show their dependence on $m_{tri}$ and $m_{coll}$. Importantly, while $h^\text{int}_{coll}$ exhibits a linear dependence, $h^\text{int}_{tri}$ shows a more complicated behavior with a negative effect from higher than linear order coefficients.
From this fact it directly follows that high order local moment correlations, $\mathcal{S}^{(4)}$, cannot stabilize the triangular state and trigger the first-order AFM-AFM transition at lower temperature, which reinforces the idea that the magnetovolume coupling is the dominant factor of this transition.

The free energy $\mathcal{G}_1$ per formula unit (three Mn's and one Pt atoms) accounting for the effect of a hydrostatic pressure $p$ is given from Eqs.\ (\ref{EQ1}) and (\ref{EQ4-MSE}) as
\begin{equation}
\begin{split}
 & \mathcal{G}_1 =\Omega_0
+f^{(2)}\left(\textbf{m}_1, \textbf{m}_2, \textbf{m}_3; \omega\right)+f^{(4)}\left(\textbf{m}_1, \textbf{m}_2, \textbf{m}_3; \omega\right) \\
 & +\frac{1}{2}V_0\gamma\omega^2+p\omega V_0-T\Big(S_1(\textbf{m}_1)+S_2(\textbf{m}_2)+S_3(\textbf{m}_3)\Big),
\label{EQ1-Pt}
\end{split}
\end{equation}
where the compressibility $\gamma$ has been taken from experiment with a value 111 GPa~\cite{doi:10.1143/JPSJ.56.4532}.
We follow the procedure explained in section \ref{Multi} and fit the internal fields $h^\text{int}_{tri}(m_{tri})$ and $h^\text{int}_{coll}(m_{coll})$ for different unit cell volumes and for the magnetic phases of interest and their dependence on volume. We find that the 8 data points in Fig.\ \ref{Mn3Pt3} for each curve can be very well fitted by
\begin{equation}
\begin{split}
& h^\text{int}_{tri}=\mathcal{S}^{(2)}_{tri}(\omega)m_{tri}+\mathcal{S}^{(4)}_{tri}m^{3}_{tri}, \\
& h^\text{int}_{coll}=\mathcal{S}^{(2)}_{coll}(\omega)m_{coll},
\label{EQ2a-Pt}
\end{split}
\end{equation}
where $\mathcal{S}^{(2)}_{tri/coll}(\omega)$ and $\mathcal{S}^{(4)}_{tri/coll}$ are compact forms of pair and quartet local moment correlations for the magnetic phases under study.
We remark that $\mathcal{S}^{(2)}_{tri}(\omega)$ and $\mathcal{S}^{(2)}_{coll}(\omega)$ are directly obtained from the volume-dependent largest eigenvalues of $\tilde{\mathcal{S}}^{(2)}_{ss'}(\textbf{q};\omega)$ at their respective values of $\textbf{q}$, i.e.\ $\mathcal{S}^{(2)}_{tri}=\max\{\tilde{u}_a(\textbf{0})\}$ and $\mathcal{S}^{(2)}_{coll}=\max\{\tilde{u}_a((0, 0, 0.5)\frac{2\pi}{a})\}$ (see section \ref{Wave}).
The fitting of Eq.\ (\ref{EQ2a-Pt}) is therefore used to extract $\mathcal{S}^{(4)}_{tri}$ only, confirming that $\mathcal{S}^{(4)}_{coll}=0$ and that higher order terms are negligible.
Both collinear and triangular states have been found to follow a good linear dependence on $\omega$ for the pairwise contribution only, $\mathcal{S}^{(2)}_{coll}=(\mathcal{S}^{(2)}_{coll,0}+\alpha_{coll}\omega)=(202-324\omega)$meV for the collinear state, and $\mathcal{S}^{(2)}_{tri}(\omega)=(\mathcal{S}^{(2)}_{tri,0}+\alpha_{tri}\omega)=(192-1086\omega)$meV and $\mathcal{S}^{(4)}_{tri}=-80$ meV for the triangular state.
From Eqs.\ (\ref{EQ1-Multi}-\ref{EQ3-Multi}) the internal energy becomes
\begin{equation}
\begin{split}
 & \langle\Omega^\text{int}\rangle_0\Big|_{\{\textbf{m}_i\}_{tri}}=\Omega_0-\frac{3}{2}\mathcal{S}^{(2)}_{tri}(\omega)m_{tri}^2 -\frac{3}{4}\mathcal{S}^{(4)}_{tri}m_{tri}^4, \\
 & \langle\Omega^\text{int}\rangle_0\Big|_{\{\textbf{m}_i\}_{coll}}=\Omega_0-\mathcal{S}^{(2)}_{coll}(\omega)m_{coll}^2,
\label{EQ3-Pt}
\end{split}
\end{equation}
where $\{\textbf{m}_i\}_{tri/coll}$ means that the local order parameters are set to describe the triangular/collinear AFM states. We stress again that whilst the triangular state has three non-zero local order parameters inside the cell, the collinear state has only two. This is due to the site with zero net magnetization, which we refer to as the third magnetically disordered sublattice, i.e.\ $m_3=0$ for the collinear AFM state.
Following the procedure explained in section \ref{MSE}, we now minimize $\mathcal{G}_1$ with respect to $\omega$ analytically and write from Eqs.\ (\ref{EQ1-Pt}) and (\ref{EQ3-Pt}) that
\begin{equation}
\begin{split}
\omega\Big|_{\{\textbf{m}_i\}_{tri}}= & \frac{\alpha_{tri}}{V_0\gamma}\frac{3}{2}m_{tri}^2-\frac{p}{\gamma}, \\
\omega\Big|_{\{\textbf{m}_i\}_{coll}}= & \frac{\alpha_{coll}}{V_0\gamma}m_{coll}^2-\frac{p}{\gamma},
\label{EQ4-Pt}
\end{split}
\end{equation}
where $\alpha_{tri}=\sum_{ij}\alpha_{tri,ij}$ and $\alpha_{coll}=\sum_{ij}\alpha_{coll,ij}$.
Substituting Eq.\ (\ref{EQ4-Pt}) into Eq.\ (\ref{EQ1-Pt}) gives
\begin{equation}
\begin{split}
 & \mathcal{G}_1\Big|_{\{\textbf{m}_i\}_{tri}}=  \\
 & \Omega_0-3\left[\frac{1}{2}\mathcal{S}^{(2)}_{tri,0}-\frac{p\alpha_{tri}}{\gamma}\right]m^2_{tri}-\frac{3}{4}\left[\mathcal{S}^{(4)}_{tri}+\frac{3\alpha_{tri}^2}{2V_0\gamma}\right]m^4_{tri} \\
 & -T\Big(S_1(m_{tri})+S_2(m_{tri})+S_3(m_{tri})\Big)-p^2V_0\gamma^{-1}, \\
 & \mathcal{G}_1\Big|_{\{\textbf{m}_i\}_{coll}}= \\
 & \Omega_0-2\left[\frac{1}{2}\mathcal{S}^{(2)}_{coll,0}-\frac{p\alpha_{coll}}{\gamma}\right]m^2_{coll}-\frac{1}{2}\frac{\alpha_{coll}^2}{V_0\gamma}m^4_{coll} \\
 & -T\Big(S_1(m_{coll})+S_2(m_{coll})+S_3(0)\Big)-p^2V_0\gamma^{-1}.
\label{EQ5-Pt}
\end{split}
\end{equation}
Hence, the magnetovolume coefficients $\{\alpha_{tri}, \alpha_{coll}\}$ give rise to fourth order contributions that in principle can trigger the first-order AFM-AFM phase transition. This is indeed what we have found in our calculations after minimizing Eq.\ (\ref{EQ5-Pt}) at $p=0$ at different temperatures. We find the first-order AFM-AFM transition to occur for lattice parameters $a>a_c=3.93$\AA, from which both transition temperatures exist with $T_{N}>T_{tr}$. Below this value the collinear AFM state is entirely suppressed and a single second-order triangular AFM-PM transition occurs. $T_{N}\approx$ 800K for $a>a_c$, somewhat higher than the experimental value of 475K~\cite{PhysRev.171.574,doi:10.1063/1.2163485,doi:10.1143/JPSJ.56.4532}. Notably, our DFT-DLM approach correctly predicts the two triangular and collinear AFM orderings as the most stable phases and explains the occurrence of the first-order transition as a magnetovolume driven effect. 

\begin{figure}[t]
\centering
\includegraphics[clip,scale=0.88]{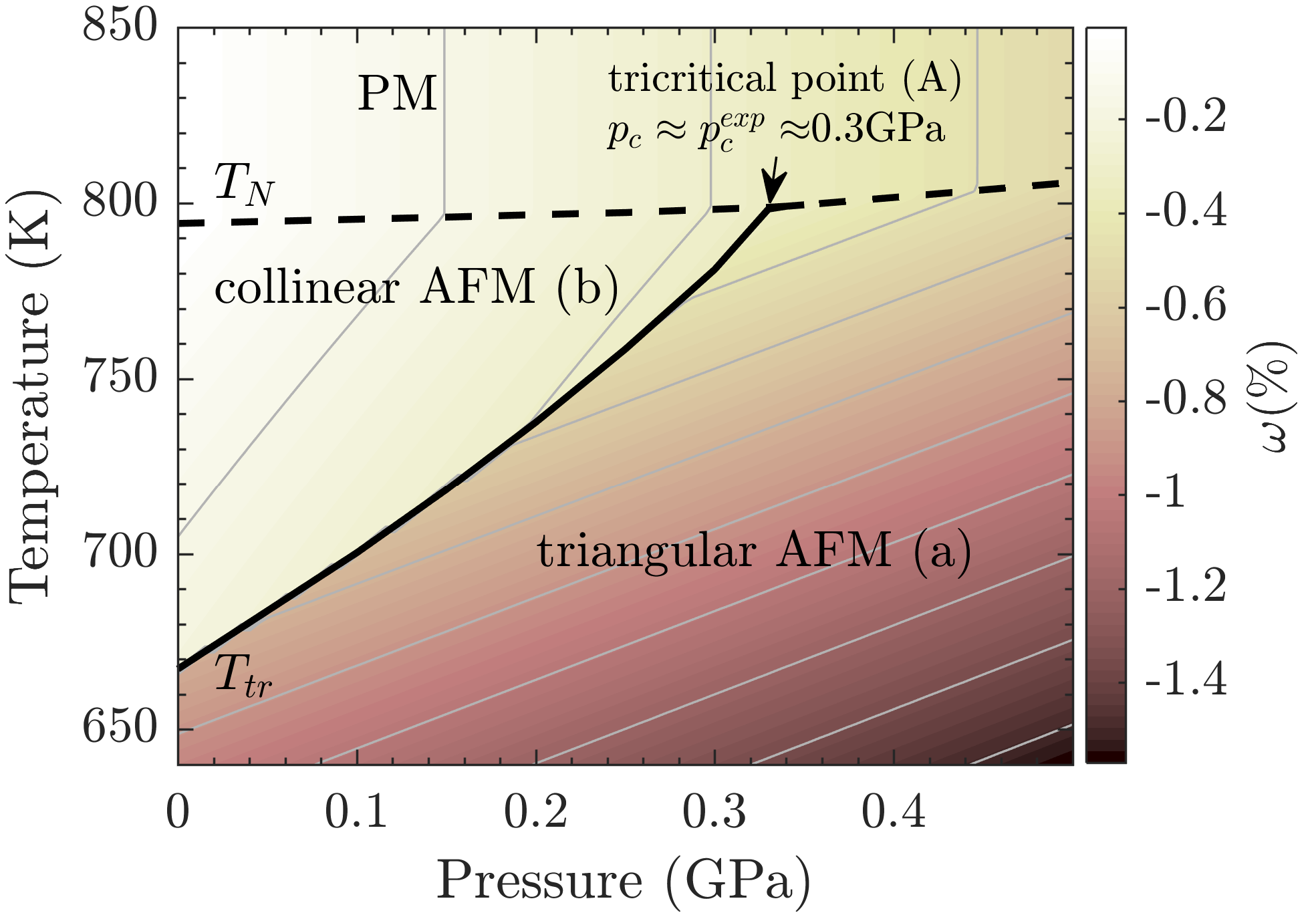}
\caption{\textit{Ab initio} pressure-temperature magnetic phase diagram of Mn$_3$Pt. The color scheme encodes the relative change of volume with respect to the PM state, $\omega$. Thick black lines indicate first-order (solid) and second-order (dashed) magnetic phase transitions and letters in brackets link to panels of Fig.\ \ref{Mn3Pt1}. A tricritical point (A) with $p_c\approx p_c^{exp}$~\cite{doi:10.1143/JPSJ.56.4532} is marked.}
\label{Mn3Pt4}
\end{figure}

To construct the pressure-temperature magnetic phase diagram we firstly choose a reference lattice parameter for the PM state that best describes the $p=0$ state, i.e.\ $a_\text{PM}=3.98$\AA\ ($V_0=a_\text{PM}^3$). This sufficiently expands the unit cell to stabilize the collinear AFM state at high $T$ and gives a temperature span for its stability of $T_{N}-T_{tri}=125$K, which agrees well with the experimental value after rescaling with our higher $T_{N}$.
We show in Fig.\ \ref{Mn3Pt4} the \textit{ab initio} magnetic phase diagram obtained.
For increasing values of $p$, $T_{tr}$ rises and eventually reaches a tricritical point (A) at $p_c=0.33$GPa, remarkably close to the experimental value~\cite{doi:10.1143/JPSJ.56.4532}. At higher values $p>p_c$ the collinear AFM order disappears at all temperatures. Our method is able to distinguish between second- and first-order phase transitions, which we indicate with continuous and dashed lines in the figure. Moreover, the volume change at $T_{tr}$ and $p=0$ is $\omega=0.6$\%, significant but somewhat below experimental findings (2.4\%).

\subsubsection{On the deficiencies of the collinear AFM state}

To describe vanishing magnetic order from fluctuating local moments on sublattices at the third sites in the collinear AFM phase, we have kept $m_3=0$ even at temperatures well below $T_{N}$. However, in the limit of approaching $T=0$ K all magnetic sites should be fully ordered at every magnetic phase, i.e.\ $\{m_n=1\}$. This means that some net magnetic order should develop at the third site at intermediate temperatures, even if the triangular state was not stabilized. This is something that was pointed out by Long~\cite{0953-8984-3-36-017}, who noted the deficiencies of such a collinear AFM state. He thus proposed a new magnetic structure compatible with a modulation of $\textbf{q}=(0, 0, 0.5)\frac{2\pi}{a}$ and with non-zero magnetic moment densities at every magnetic site. To study this effect we have calculated the local magnetic fields for magnetic phases in which only the third magnetic site has non-zero magnetic ordering, $\{m_1=m_2=0, m_3\neq0\}$, in both FM ($\textbf{q}=\textbf{0}$) and collinear AFM ($\textbf{q}=(0, 0, 0.5)\frac{2\pi}{a}$) modulations. From this calculation we can investigate at what temperature this site begins to develop net spin polarization when embedded in an otherwise PM environment, i.e.\ we study the effect from the leading second-order coefficients in the free energy. In both cases we find that the electronic interactions set a non-zero magnetic order at the third site only for temperatures around 100K or below. These values are very small compared with the transition temperatures and so strongly indicate that the collinear AFM state as described in Fig.\ \ref{Mn3Pt1}(b) should persist down to $T_{tr}$. In addition, we also show that the magnetovolume coupling produces the leading quartic coefficients in the interacting part of free energy $\mathcal{G}_1$ (Eq.\ (\ref{EQ1}). Thus, as the temperature decreases the triangular AFM phase stabilizes before $m_3\neq 0$ occurs for the collinear AFM state. As suggested by Hirano \textit{et al}.~\cite{HIRANO19951975}, however, further measurements are necessary to corroborate the nature of the collinear AFM phase.
We finally remark that our analysis adds to the findings from Kota \textit{et al}.'s DFT calculations~\cite{Kota1}.

\subsubsection{The barocaloric effect of Mn$_3$Pt}

We finish this section by evaluating the BCE in Mn$_3$Pt to demonstrate how caloric effects are accessible from the theory. We follow a similar procedure as performed in our work for the elastocaloric effect in Mn$_3$GaN~\cite{PhysRevB.95.184438}.
After inspecting Fig.\ \ref{Mn3Pt4} clearly the largest thermodynamic changes occur around the first-order AFM-AFM phase transition $T_{tr}$ at $p=0$. For increasing values of $p$ the transition loses its first-order character and so the associated entropy change becomes smaller. We calculate the total entropy difference above and below $T_{tr}$ using Eq.\ (\ref{EQ3a-Cal}), as well as Eqs.\ (\ref{EQ10-Theo}) and (\ref{EQ1-Cal}). For the corresponding phases the local order parameters are $m_{coll}=0.495$ and $m_{tri}=0.431$, respectively. From this it follows that the magnetic contribution is small, $\Delta S_\text{mag}=2$J/kgK. Moreover, $\Delta S_{elec}$ is negligible owing to the tiny change of the density of states at the Fermi energy between these two phases. Despite adiabatic temperature changes being potentially large due to the large value of $\text{d}T_{tr}/\text{d}p$, the isothermal entropy change is small meaning that Mn$_3$Pt is not a good candidate for magnetic cooling based on hydrostatic pressure application, unlike the Mn-based antiperovskites~\cite{PhysRevB.95.184438,PhysRevX.8.041035}.

\subsection{Mn$_3$A (A=Sn, Ga, Ge): hexagonal and tetragonal structures}
\label{NTMSect}

\subsubsection{Experimentally observed properties and published theoretical calculations}

\begin{figure}[t]
\centering
\includegraphics[clip,scale=0.13]{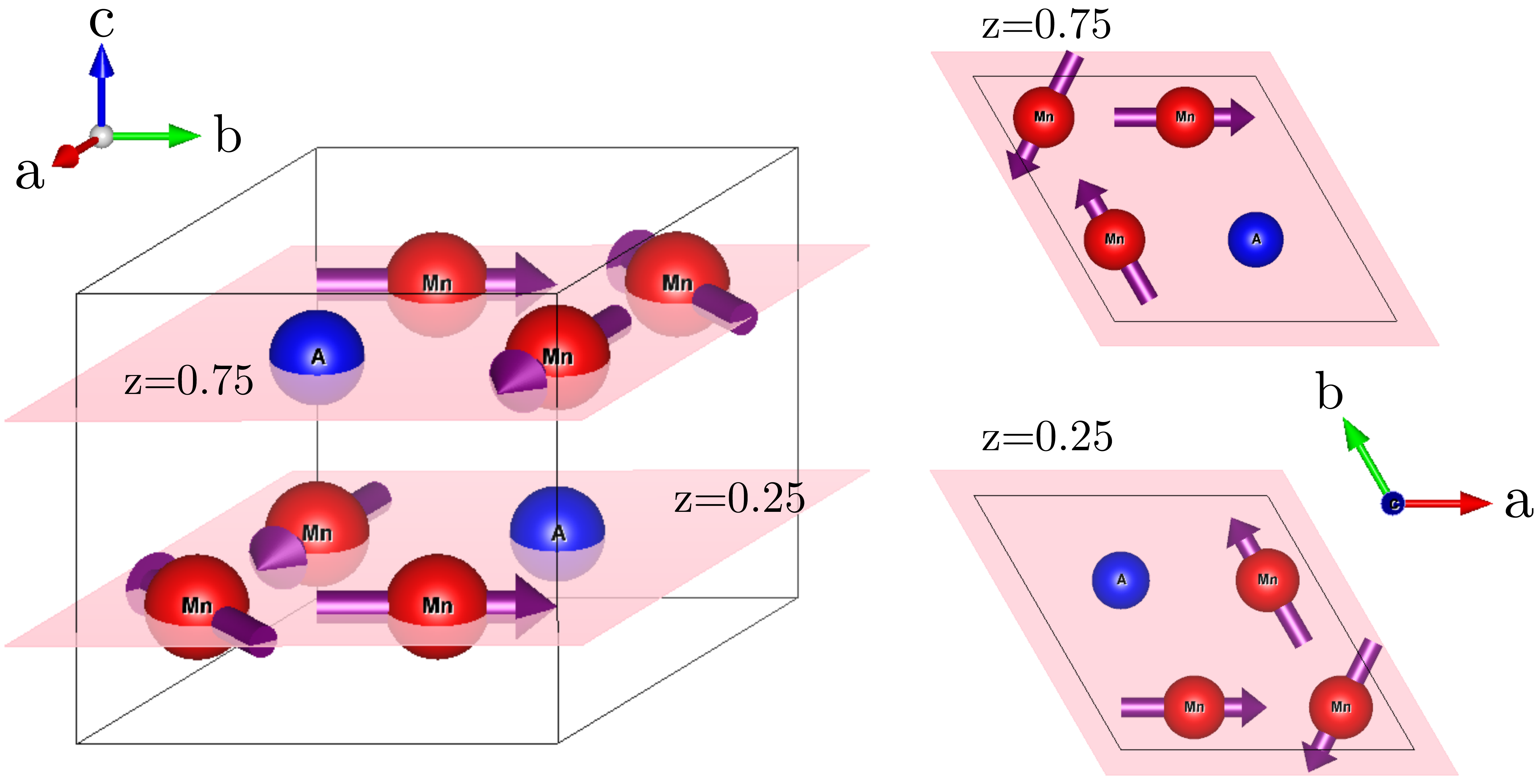}
\caption{Hexagonal magnetic structure of Mn$_3$A (A=Sn, Ga, Ge). Magenta arrows are along the local magnetic order parameters $\{\textbf{m}_n\}$, and  z indicates the height with respect to the $\hat{c}$ lattice direction. The figure shows the structure for the ideal case of a fully compensated triangular state.}
\label{FigHexagonal}
\end{figure}

At high temperatures Mn$_3$A (A=Sn, Ga, Ge) can crystallize into the hexagonal $DO_{19}$-type structure shown in Fig.\ \ref{FigHexagonal}~\cite{doi:10.1063/1.3699489,0953-8984-2-47-015,doi:10.1143/JPSJ.51.2478,KREN19701653,Mn3GeINT,YAMADA1988311,0022-3727-47-30-305001}. Equidistant layers containing both Mn and A atoms, preserving the formula unit proportion, are stacked perpendicular to the $c$ axis. Owing to the presence of geometrically frustrated AFM interactions, a triangular AFM order, as illustrated by the arrows in the figure, forms in experiment below a N\'eel temperature that spans around 400K depending on the choice of A, as indicated in table \ref{TabHex}~\cite{doi:10.1063/1.3699489,0953-8984-2-47-015,KREN19701653,Mn3GeINT,YAMADA1988311,0022-3727-47-30-305001}. Much interest for these systems is driven by the fact that a strong anomalous Hall effect (AHE) has been experimentally observed for all three materials (A=Sn, Ga, Ge)~\cite{Nakatsuji1,Mn3GaHallEffect,Nayake1501870}, as well as by the demonstration from purely symmetry arguments that spin-polarized currents can be induced owing to the non-collinearity~\cite{PhysRevLett.119.187204}. Contrary to chiral FM structures with a non-zero out-of-plane component, in which the AHE can be driven by a real-space topological effect~\cite{PhysRevB.62.R6065,Taguchi1}, in co-planar chiral antiferromagnets with local magnetic moments lying within the basal plane the AHE is generated by the spin-orbit coupling (SOC)~\cite{PhysRevLett.112.017205}. The hexagonal family Mn$_3$A shows a SOC that distorts the triangular arrangement, generating a weak non-zero net magnetization~\cite{doi:10.1143/JPSJ.51.2478,0953-8984-2-47-015,doi:10.1063/1.4929447,NAGAMIYA1982385}.
These systems have been in consequence focus of extensive theoretical work based on $T=0$K \textit{ab initio} calculations in order to evaluate their potential to induce large Hall effects and study possible triangular configurations~\cite{PhysRevB.95.075128,PhysRevLett.112.017205,0295-5075-108-6-67001,PhysRevLett.76.4963}.

Moreover, on annealing for long periods of time the hexagonal Mn$_3$Ga and Mn$_3$Ge phases, at low temperature they transform into the tetragonal $DO_{22}$ phase shown in Fig.\ \ref{Mn3Ga1_tog}(a)~\cite{KREN19701653,Mn3GeINT,PhysRevB.77.054406,0953-8984-25-20-206006}. This structure contains two non-equivalent atomic positions denoted as 2$b$ and 2$d$. The magnetic phase that stabilizes is in sharp contrast with the triangular AFM state observed in its hexagonal counterpart. It is a ferrimagnetic (FIM) state in which in a unit cell two of the Mn magnetic local order parameters point in one direction and the remaining along the opposite. These directions are indicated by magenta and blue arrows in the figure and are associated with the 4$d$ and 2$b$ positions, respectively. The magnetism here is collinear so that spintronic properties from non-collinear and chiral AFM are not present~\cite{PhysRevB.95.075128}. However, the FIM phase has a high Curie temperature and a low net magnetic moment. This together with the fact that epitaxial films can be grown to achieve high perpendicular magnetic anisotropy have made these systems very appealing for spin transfer torque applications set in small and thermally stable designs~\cite{doi:10.1063/1.2722206,PhysRevB.83.020405,doi:10.1063/1.4754123,doi:10.1063/1.5013667,YOU201740,GUTIERREZPEREZ201820,BANG201663}.

A consistent temperature-dependent study of hexagonal and tetragonal Mn$_3$A, although highly desirable, is still missing. In this section we address this by applying  our DLM picture without SOC and so study the consequences of thermal fluctuations on magnetic frustration and magnetic phase stability. The results obtained can be used as a starting point for subsequent studies which include relativistic effects and which are pitched at uncovering subtle SOC effects on electronic and  magnetic structure at finite temperatures.

\subsubsection{Application of DLM theory to hexagonal Mn$_3$Sn, Mn$_3$Ga and Mn$_3$Ge}
\label{hexa}

\begin{table*}
\begin{center}
\begin{tabular}{ c c c c | c c c c c c }
 \hline\hline
 & Ref.\ & $a=b$ (\AA) & $c$ (\AA) & $T_{N}^\text{exp}$ (K) & $T_{N}^\text{theo}$ (K) & $\mu_\text{Mn}$ ($\mu_\text{B}$) & $\mathcal{S}^{(2)}_{tri}$ (meV) & $\mathcal{S}^{(4)}_{tri}$ (meV) & $\mathcal{S}^{(6)}_{tri}$ (meV) \\
 \hline
Mn$_3$Sn & ~\cite{doi:10.1063/1.3699489,0953-8984-2-47-015}      & 5.66 & 4.53 & 420 & 495 & 2.98 & 131.2 & -80$\pm$2 & 59$\pm$2 \\
Mn$_3$Ga & ~\cite{KREN19701653}                                  & 5.40 & 4.35 & 470 & 586 & 2.65 & 152.0 & -61.5$\pm$0.4 & 19.5$\pm$0.4 \\
Mn$_3$Ge & ~\cite{Mn3GeINT,YAMADA1988311,0022-3727-47-30-305001} & 5.36 & 4.32 & 365 & 486 & 2.43 & 127.4 & -68$\pm$2 & 46$\pm$4 \\
 \hline\hline
\end{tabular}
\caption{Application of the theory to hexagonal Mn$_3$A (A=Sn, Ga, Ge). The table shows the lattice parameters taken directly from experiment, and results for local moment sizes, direct pair $\mathcal{S}^{(2)}_{tri}$ and higher order local moment correlations, $\mathcal{S}^{(4)}_{tri}$ and $\mathcal{S}^{(6)}_{tri}$, and N\'eel transition temperatures, $T_N^\text{theo}$. The latter are compared with experimental values, $T_N^\text{exp}$. The given errors are extracted from a least squares fitting of the \textit{ab-initio} calculations.}
\label{TabHex}
\end{center}
\end{table*}

\begin{figure*}[t!]
\centering
\includegraphics[clip,scale=0.72]{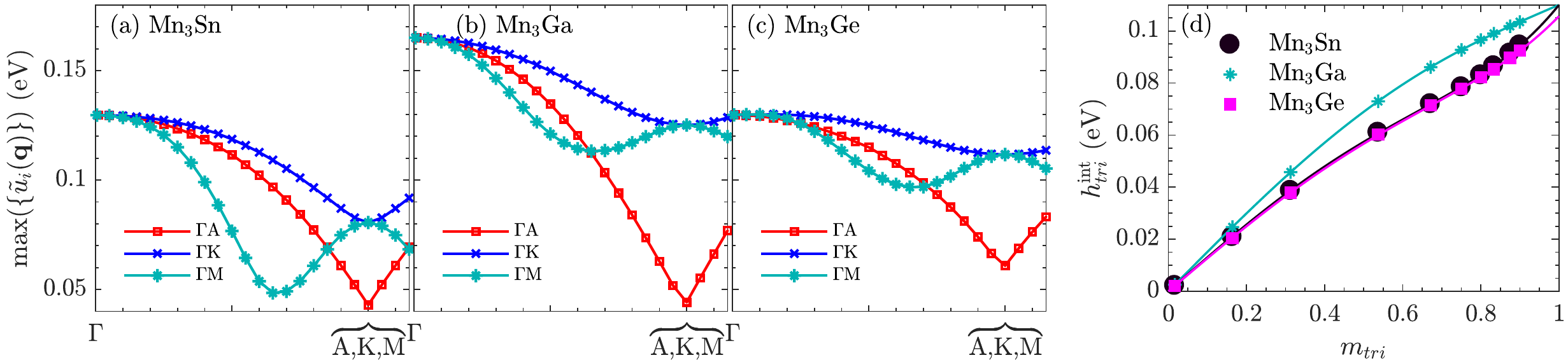}
\caption{Dependence of the largest eigenvalue of $\tilde{\mathcal{S}}^{(2)}_{ss'}(\textbf{q})$ for hexagonal (a) Mn$_3$Sn, (b) Mn$_3$Ge, and (c) Mn$_3$Ga. The results are given for three characteristic directions within the Brillouin zone through the $\Gamma$ point. (d) Strength of the local field against the local order parameter for the three studied materials, Mn$_3$Sn (black circles), Mn$_3$Ga (light blue stars), and Mn$_3$Ge (magenta squares). Continuous lines are the fitted functions of Eq.\ (\ref{EQ1-Hex}) using the data points.}
\label{figHexWave}
\end{figure*}

The first step is to calculate the direct correlation function $\tilde{\mathcal{S}}^{(2)}_{ss'}(\textbf{q})$ and obtain the most stable magnetic phases. For the three hexagonal materials Mn$_3$Sn, Mn$_3$Ga and Mn$_3$Ge under study local magnetic moments with similar sizes established at each Mn site. In table \ref{TabHex} we give these values as well as the lattice parameters used, taken from experiment.
Since the crystallographic unit cell of the hexagonal Mn$_3$A contains six magnetic positions, $\tilde{\mathcal{S}}^{(2)}_{ss'}(\textbf{q})$ is a 6$\times$6 matrix.
Fig.\ \ref{figHexWave}(a,b,c) shows the largest eigenvalue of $\tilde{\mathcal{S}}^{(2)}_{ss'}(\textbf{q})$, $\max\{\tilde{u}_a(\textbf{q})\}$, corresponding to the most stable magnetic phase, against the wave vector inside the Brillouin zone. For the three materials it peaks at $\textbf{q}=\textbf{0}$. This means that the magnetic unit cell matches the crystallographic unit cell and so there are no rotations of the local moment directions from one cell to another. An inspection of the eigenvector components gives the relative orientations between the six local order parameters inside one magnetic unit cell. For the three systems these components perfectly match cosines of angles describing a fully compensated triangular state as shown in Fig.\ \ref{FigHexagonal}. The eigenvector is for example $(1,-\frac{1}{2},-\frac{1}{2},1,-\frac{1}{2},-\frac{1}{2})$, where the first three components correspond to Mn atoms at z=0.25 and the next three at z=0.75. After examining the $\textbf{q}$-dependence of $\tilde{\mathcal{S}}^{(2)}_{ss'}(\textbf{q})$ thoroughly we did not find signatures of $\textbf{q}\neq\textbf{0}$ phases that would be more stable, i.e.\ the magnetic correlations that are precursors of the triangular state are the strongest.
Importantly, in striking contrast to cubic Mn$_3$A (A=Pt, Ir, Rh) here we found that the eigenvalue of $\tilde{\mathcal{S}}^{(2)}_{ss'}(\textbf{q})$ at $\textbf{q}=\textbf{0}$ is consistently the largest for different unit cell volumes, and therefore magnetoelastic effects are not relevant.

We now model the triangular AFM state at lower temperatures. This firstly requires that this magnetic state is described in terms of the local order parameters $\{\textbf{m}_n\}$, which describe the magnetic order of the triangular AFM phase in Fig.\ \ref{FigHexagonal}. Since this state is fully compensated, the six vectors $\{\textbf{m}_n\}$ in the unit cell have the same length and only differentiate by in-plane rotations of 120 degrees. Lowering $T$ does not change their orientations and induces an identical increase of their magnitudes only. Hence, the entire magnetic state is characterized by one common local order parameter size that we label as $m_{tri}$. Evidently, the effective fields $\{\textbf{h}^\text{int}_n\}$, given in Eq.\ (\ref{EQ14-Theo}), have the same length $h^\text{int}_{tri}$ at each Mn site too. In Fig.\ \ref{figHexWave}(d) we show the DFT-DLM calculation of $h^\text{int}_{tri}$ as a function of $m_{tri}$ for A=Sn, Ga, and Ge. Although small, there is a clear deviation from a linear dependence for increasing $m_{tri}$, which explicitly indicates that higher order local moment correlation effects are significant (see section \ref{Multi}). Indeed, we have found that the local field data is very well fitted for every choice of atom A by
\begin{equation}
h^\text{int}_{tri}=\mathcal{S}^{(2)}_{tri}m_{tri}+\mathcal{S}^{(4)}_{tri}m_{tri}^3+\mathcal{S}^{(6)}_{tri}m_{tri}^5,
\label{EQ1-Hex}
\end{equation}
where $\mathcal{S}^{(2)}_{tri}$ is directly determined from the largest eigenvalue of $\tilde{\mathcal{S}}^{(2)}_{ss'}(\textbf{q})$, which corresponds to the triangular AFM state, as $\mathcal{S}^{(2)}_{tri}=\max\{\tilde{u}_a(\textbf{q}=\textbf{0})\}$.
We remark that we have used 10 data points, shown as single points in Fig.\ \ref{figHexWave}(d), to fit the two constants $\mathcal{S}^{(4)}_{tri}$ and $\mathcal{S}^{(6)}_{tri}$. Their values are tabulated in table \ref{TabHex}, and the fitted functions are shown in Fig.\ \ref{figHexWave}(d) as continuous lines.

As explained in section \ref{Multi}, $\mathcal{S}^{(2)}_{tri}$ and $\{\mathcal{S}^{(4)}_{tri}, \mathcal{S}^{(6)}_{tri}\}$ are, respectively, compact forms of direct local moment correlations for the magnetic state under study, which is the triangular AFM state in this case.
$\mathcal{S}^{(4)}_{tri}$ and $\mathcal{S}^{(6)}_{tri}$  describe the feedback between the spin-polarized electronic structure and the increase of magnetic order. Their calculation from the data of Fig.\ \ref{figHexWave}(d) directly provides the magnetic material's free energy from Eqs.\ (\ref{EQ1}), (\ref{EQ14-Theo}), and (\ref{EQ1-Multi}),
\begin{equation}
\mathcal{G}_1=\Omega_0-\frac{1}{2}\mathcal{S}^{(2)}_{tri}m_{tri}^2-\frac{1}{4}\mathcal{S}^{(4)}_{tri}m_{tri}^4-\frac{1}{6}\mathcal{S}^{(6)}_{tri}m_{tri}^6-TS_n(m_{tri}),
\label{EQ2-Hex}
\end{equation}
per magnetic site $n$. Minimizing Eq.\ (\ref{EQ2-Hex}) with respect to $m_{tri}$ yields $m_{tri}$ itself as a function of $T$.
The values of $\mathcal{S}^{(4)}_{tri}$ and $\mathcal{S}^{(6)}_{tri}$ obtained are negative or small so that the triangular AFM-PM transition at $T_N^\text{theo}$ has been found to be of second order, in agreement with experiment.
Remarkably, the theory correctly captures trends and magnitudes of $T_N$ for different choices of atom A in line with experimental findings as shown in table \ref{TabHex}.
Interestingly, from Fig.\ \ref{figHexWave}(d) we observe that Mn$_3$Sn and Mn$_3$Ge have a similar behavior in qualitatively significant contrast compared with Mn$_3$Ga. The main differences are the higher gradient of $h^\text{int}_{tri}$ at small values of $m_{tri}$ for Mn$_3$Ga, leading to the larger values of both $T_N^\text{theo}$ and $\mathcal{S}^{(2)}_{tri}$, as well as the different concavity around $m_{tri}\approx0.8$ as a direct consequence of the presence of $\mathcal{S}^{(4)}_{tri}$ and $\mathcal{S}^{(6)}_{tri}$.
These higher order components are significant in hexagonal Mn$_3$A, and with the consequence that any effective pairwise interactions between local moments calculated for different magnetic states will adopt different values.

\subsubsection{Application of DLM theory to tetragonal Mn$_3$Ga and Mn$_3$Ge}
\label{tetra}

\begin{table*}[t]\centering
\begin{tabular}{cccccc ccc ccccc}
\hline\hline
\multicolumn{6}{l}{FIM state in tetragonal Mn$_3$A} & \multicolumn{3}{c}{quadratic terms (meV)} & \multicolumn{5}{c}{higher order terms (meV)} \\
\cmidrule(rl){7-9}\cmidrule(rl){10-14}
A & $a$ (\AA) & $c$ (\AA) & $\mu_1$ & $\mu_{2}$=$\mu_{3}$ & $T_{tr}^\text{A}$ (K) & $\mathcal{S}^{(2)}_1$ & $\mathcal{S}^{(2)}_2$ & $\mathcal{S}^{(2)}_3$ & $\mathcal{S}^{(4)}_1$ & $\mathcal{S}^{(6)}_1$ & $\mathcal{S}^{(6)}_2$ & $\mathcal{S}^{(6)}_3$ & $\mathcal{S}^{(6)}_4$ \\ 
    \hline
Ga & 3.91 & 7.10 & 2.32$\mu_\text{B}$ & 2.02$\mu_\text{B}$ & 995 & 9.60  & 118 & -131.5 & $\approx$0   & 54$\pm$2     & 47$\pm$2     & 16.0$\pm$0.6 & -10.8$\pm$0.7 \\
Ge & 3.81 & 7.26 & 2.19$\mu_\text{B}$ & 1.56$\mu_\text{B}$ & 814 & 12.7 & 110 & -99.56 & -27.6$\pm$0.8 & 36.3$\pm$0.9 & 35.6$\pm$0.8 & 11.0$\pm$0.3 & -7.9$\pm$0.3 \\
\hline\hline
\end{tabular}
\caption{Quadratic and higher order terms as given in Eq.\ (\ref{EQ2-Ga}) for the FIM state of tetragonal Mn$_3$Ga and  Mn$_3$Ge. The lattice parameters are taken from experiment~\cite{KREN19701653,PhysRevB.77.054406,Mn3GeINT,doi:10.1063/1.4754123}. The given errors are extracted from a least squares fitting of the \textit{ab-initio} data. The table also shows the magnetic moment sizes and transition temperatures obtained.}
\label{TabFIM}
\end{table*}

\begin{figure*}[t]
\centering
\subfigure{\label{fig:b}\includegraphics[clip,scale=0.89]{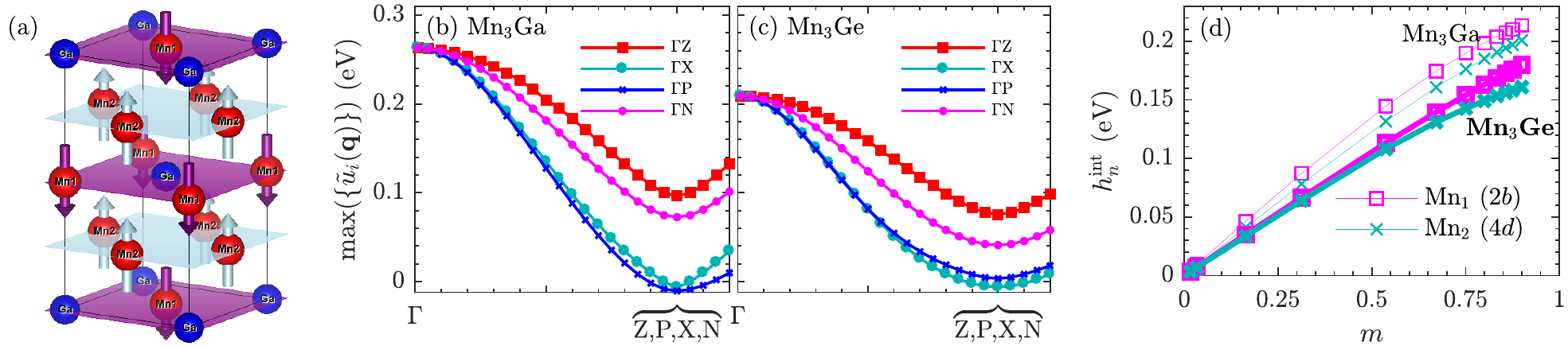}}
\caption{(a) Ferrimagnetic unit cell of tetragonal Mn$_3$Ga and Mn$_3$Ge. Magenta and blue arrows are used to indicate "up" and "down" local order parameter orientations.
(b-c) Wave vector dependence of the largest eigenvalue of $\tilde{\mathcal{S}}^{(2)}_{ss'}(\textbf{q})$ for Mn$_3$Ga and Mn$_3$Ge. Results are shown for four characteristic directions from the $\Gamma$ point to special points on the Brillouin zone boundary. (d) Absolute values of the local fields for the two non-equivalent sites, associated with Mn$_1$ and Mn$_2$(Mn$_3)$, against the local order parameters when $\textbf{m}=\textbf{m}_1=\textbf{m}_2=-\textbf{m}_3$. Lighter and wider lines are used for Mn$_3$Ga and Mn$_3$Ge, respectively.}
\label{Mn3Ga1_tog}
\end{figure*}

The tetragonal structure shown in Fig.\ \ref{Mn3Ga1_tog}(a) can be equivalently described as a bcc lattice with four atoms per unit cell. One Ga atom is positioned at (0, 0, 0) and three Mn atoms at $\textbf{r}_{\text{Mn}_1}$=(0, 0, 1/2), $\textbf{r}_{\text{Mn}_2}$=(1/2, 0, 1/4), and $\textbf{r}_{\text{Mn}_3}$=(0, 1/2, 1/4). The last two atomic positions are equivalent and associated with the 4$d$ site, while the first Mn position corresponds to the 2$b$ site. Similarly to the Cu$_3$Au-type case in Mn$_3$Pt, the direct correlation function $\tilde{\mathcal{S}}_{ss'}^{(2)}(\textbf{q})$ in the bcc basis is a 3$\times$3 matrix with three eigenfunctions $\{\tilde{u}_i(\textbf{q}), i=1,2,3\}$. In Fig.\ \ref{Mn3Ga1_tog}(b) and \ref{Mn3Ga1_tog}(c) we show the largest value among these functions for relevant directions inside the Brillouin zone.
We remark that we consistently obtained magnetic moment sizes $\mu_1>\mu_2=\mu_3$, as observed experimentally~\cite{KREN19701653,doi:10.1063/1.4754123} and in agreement with other first principles calculations~\cite{doi:10.1063/1.4970691}. From Fig.\ \ref{Mn3Ga1_tog}(b-c) we can see that the peak is at $\textbf{q}=\textbf{0}$ and that there are no other competing modulations. From Eq.\ (\ref{EQ12-Wave}) we calculate the FIM-PM transition temperature to be $T_{tr}^\text{Ga}=995$K and $T_{tr}^\text{Ge}=814$K. Since in experiment the material transforms into the hexagonal structure with increase in temperature before going through $T_{tr}$, our results cannot be directly compared. Our data is in agreement with literature, however, because the structural transition occurs below our transition temperature predictions for Mn$_3$Ga~\cite{KREN19701653} and the extrapolated value for Mn$_3$Ge, 920K~\cite{doi:10.1063/1.4754123}.

The eigenvector components obtained for the maximum eigenvalue after diagonalizing $\tilde{S}_{ss'}^{(2)}(\textbf{0})$ are (-0.614, +0.558, +0.558) and (-0.587, +0.573, +0.573), for Mn$_3$Ga and Mn$_3$Ge respectively. These describe the FIM phase with Mn atoms at 4$d$ sites developing identical local order parameters being antiparallel to the atom at position 2$b$. The different values among the eigenvector components indicate that Mn$_1$ develops an average local magnetization higher than Mn$_2$ and Mn$_3$, i.e.\ $m_1>m_2=m_3$. This becomes more evident after studying how the local fields behave if identical amount of local magnetic ordering is imposed at all sites, which we prescribe by a quantity labeled as $m$, that is $\textbf{m}=\textbf{m}_2=\textbf{m}_3=-\textbf{m}_3$. Indeed, Fig.\ \ref{Mn3Ga1_tog}(d) shows that $h^\text{int}_n(m)$ at Mn$_1$ is slightly larger compared with Mn$_{2(3)}$.

We fit the local field DFT-DLM data, $\{\textbf{h}^\text{int}_n\}$, in order to extract the first derivatives of the internal magnetic energy $\langle\Omega^\text{int}\rangle_0$ with respect to the local order parameters via Eq.\ (\ref{EQ14-Theo}). We have used the data points shown in Fig.\ \ref{Mn3Ga1_tog}(d) as well as many other magnetic configurations comprising $\{\textbf{m}_1=-m_1\hat{z}, \textbf{m}_2=\textbf{m}_3=\textbf{0}\}$, $\{\textbf{m}_1=\textbf{0}, \textbf{m}_2=\textbf{m}_3=m_2\hat{z}\}$, and $\{\textbf{m}_1=-m_1\hat{z}, \textbf{m}_2=\textbf{m}_3=m_2\hat{z}\}$, for mixed magnetic orderings ranging as $m_1$(and/or $m_2$)=0.05, 0.1, 0.5, 1, 2, 3, 4, 5, 6, 7, 8, and 10, which thoroughly samples the $\{\textbf{m}_n\}$-space for the FIM state. In total there were 216 data points to fit, which we found to be very well described by just five  correlation function quantities in
\begin{equation}
\begin{split}
 & \langle\Omega^\text{int}\rangle_0=-\frac{1}{2}\mathcal{S}^{(2)}_{1}m_1^2-\frac{1}{2}\mathcal{S}^{(2)}_2(m_2^2+m_3^2) \\
 & -\mathcal{S}^{(2)}_3\textbf{m}_1\cdot(\textbf{m}_2+\textbf{m}_3)-\frac{1}{4}\mathcal{S}^{(4)}_1(m_2^4+m_3^4) \\
 & -\mathcal{S}^{(6)}_1\textbf{m}_1^3\cdot(\textbf{m}_2^3+\textbf{m}_3^3)-\mathcal{S}^{(6)}_2 m_1^2(m_2^4+m_3^4) \\
 & -\mathcal{S}^{(6)}_3\textbf{m}_1\cdot(\textbf{m}_2^5+\textbf{m}_3^5)-\mathcal{S}^{(6)}_4\textbf{m}_1^5\cdot(\textbf{m}_2+\textbf{m}_3).
\label{EQ2-Ga}
\end{split}
\end{equation}
The quadratic terms $\{\mathcal{S}^{(2)}_1, \mathcal{S}^{(2)}_2, \mathcal{S}^{(2)}_3\}$ are directly extracted from the calculation of the direct correlation function $\tilde{\mathcal{S}}^{(2)}_{ss'}(\textbf{q})$. The higher order terms, compactly contained in $\{\mathcal{S}^{(4)}_1, \mathcal{S}^{(6)}_1, \mathcal{S}^{(6)}_2, \mathcal{S}^{(6)}_3, \mathcal{S}^{(6)}_4\}$, are only important to numerically capture small deviations of a nearly linear dependence of $\{h^\text{int}_n\}$, as it can be seen from Fig.\ \ref{Mn3Ga1_tog}(d). The  multi-site local moment interactions that are inferred from these quantities 
for tetragonal Mn$_3$A, as also proposed by Khmelevskyi \textit{et al.}~\cite{PhysRevB.93.184404}, have only a small effect on the FIM phase and in consequence the FIM-PM phase transition is predicted to be of second-order. We show the results in table \ref{TabFIM}. The free energy expressions and  Eq.\ (\ref{EQ2-Ga}) for both Mn$_3$Ga and Mn$_3$Ge are identical in form and the sizes of constants are also similar. In particular $\mathcal{S}^{(2)}_1$ is very small compared with $\mathcal{S}^{(2)}_2$ and $\mathcal{S}^{(2)}_3$. This means that the net magnetic polarization at site 2$b$ arises as a slave effect from the stronger and leading local magnetic ordering at sites 4$d$. This is underpinned by the large value of $\mathcal{S}^{(2)}_3$, whose negative sign captures the AFM nature among interactions between both sites.
We also remark that the higher transition temperature for A=Ga is a direct consequence of the larger values of $\mathcal{S}^{(2)}_2$ and $\mathcal{S}^{(2)}_3$, which links to the temperature from which 4$d$ begins to develop a net local magnetic order.

We finish this section with a discussion of geometrically frustrated magnetism. To this end we analysed in more detail the direct local moment-local moment correlation function $\tilde{\mathcal{S}}^{(2)}_{ss'}(\textbf{q})$. We found that parallel alignment of $\textbf{m}_2$ and $\textbf{m}_3$ is energetically preferable and compatible with a picture of magnetic interactions that do not show frustration effects. We therefore concluded that tetragonal Mn$_3$A should not be regarded as a magnetically frustrated system. Our findings are in agreement with a major theoretical work on pairwise interactions carried out by Khmelevskyi \textit{et al.}~\cite{PhysRevB.93.184404}.

\section{Summary, conclusions and outlook}
\label{Conc}

We have presented an \textit{ab initio} theory for the Gibbs free energy of a magnetic material containing direct pair- and higher order local magnetic moment correlation functions, and their dependence on the lattice structure, from magnetically constrained DFT calculations.
The essence of our approach is to describe the dependence of the electronic structure on the magnetic order and lattice deformations, as well as to evaluate the impact of the electronic response has on the magnetism itself.

Potential stable magnetic structures at lower temperatures are identified by studying the instabilities of the PM state. We achieve this from the lattice Fourier transform of the direct local moment-local moment correlation function $\tilde{S}^{(2)}_{ss'}(\textbf{q})$ and its eigenvalues and eigenvectors. As an important contribution, the formalism presented in this paper has been extended to study materials with complex multi-atom unit cells and long-period magnetic structures.
A central step of our theory consists in the calculation of internal local magnetic fields $\{\textbf{h}^\text{int}_n\}$ sustaining the local moments as function of magnetic order $\{\textbf{m}_n\}$, which we use to extract the effects of higher order correlations among the local moments.
Magnetic phase diagrams containing first- and second-order magnetic phase transitions and tricritical points for temperature, magnetic field and crystal structure are obtainable. Importantly, we quantify the contribution to terms in the free energy which are fourth order with respect to the local moment magnetic order parameters from both a magnetoelastic effect as well as from a purely electronic source.
The method is also designed to calculate isothermal entropy changes and adiabatic temperature changes, which we use to evaluate caloric effects at magnetic phase transitions and hence the best refrigerating performance of a magnetic material from its phase diagram's features.

We have studied magnetic materials in the family Mn$_3$A, currently attracting much interest for spintronic applications. Their rich spectrum of magnetic phases, including non-collinear and collinear AFM and ferrimagnetic states, sets a challenging test for our \textit{ab initio} theory to explain their temperature-dependent properties and magnetic frustration.
For all cubic, hexagonal and tetragonal crystal structures we firstly calculated $\tilde{S}^{(2)}_{ss'}(\textbf{q})$. Our calculations are in excellent agreement with experiment: we predict the stabilization of the triangular AFM state for the cubic and hexagonal lattices, the ferrimagnetic state for the tetragonal lattice, as well as a collinear AFM phase for the cubic Mn$_3$Pt, which competes in stabilization with the triangular AFM state.
The values and trends of transition temperatures obtained are in very good agreement.
We then performed many calculations of $\{\textbf{h}^\text{int}_n\}$ to follow the development of magnetic order at lower temperature, captured the effect of multi-site local moment interactions for all systems from higher order local moment correlation functions, and produced phase diagrams.

The most striking results reported are for Mn$_3$A with cubic structure.
Results of $\tilde{S}^{(2)}_{ss'}(\textbf{q})$ obtained for cubic Mn$_3$Rh and Mn$_3$Ir are in stark contrast compared with those for Mn$_3$Pt. The theory directly links the instability of the collinear AFM phase to the pair correlation functions, which strongly favor the triangular AFM state in Mn$_3$Rh and Mn$_3$Ir and produce similar stability of both collinear and triangular AFM states in Mn$_3$Pt.
Notably, we have provided the origin of the first-order phase transition between the triangular AFM and collinear AFM states in Mn$_3$Pt as a magnetovolume driven effect, and its absence in Mn$_3$Ir and Mn$_3$Rh.
We constructed the temperature-pressure magnetic phase diagram of Mn$_3$Pt and obtained the same features as in experiment: compression destroys the collinear AFM order and eventually a single second-order PM-triangular AFM phase transition is observed after crossing a tricritical point. 
We also predict that the collinear AFM state is stable to low temperatures, hence ruling out the necessity to invoke other non-frustrated magnetic phases as suggested by Long~\cite{0953-8984-3-36-017}.
In addition, by calculating the isothermal entropy change at the AFM-AFM transition and inspecting the magnetic phase diagram's features we showed that Mn$_3$Pt is not a good barocaloric material despite its strong first-order transition.

From this extensive study for all Mn$_3$A materials we have shown that  the effect of multi-site local moment interactions are not significant in tetragonal Mn$_3$A and so theoretical approaches based on effective pairwise interactions are sufficient. However, for the cubic and hexagonal structures higher order terms are non-negligible in the triangular AFM state, and so calculations going beyond a simple Heisenberg picture must be considered for an accurate description at finite temperature and different magnetic states.
We also showed that leading magnetic interactions giving rise to the ferrimagnetic state in tetragonal Mn$_3$A are in overall satisfied and so this system should not be regarded as a frustrated magnetic material.

Finally, we stress that our theory can be used within a fully relativistic scheme so that spin-orbit coupling (SOC) effects can be taken into account and spin-dependent transport can be modeled at different temperatures and states of magnetic order~\cite{PhysRevB.95.155139}. Although our results without SOC capture the major aspects of the magnetism of Mn$_3$A, the effect of the multi-site local moment interactions could be crucial to describe temperature-dependent properties and possible magnetic phase transitions when relativistic effects are taken into account, such as the weak FM component observed in the triangular AFM state for the hexagonal structure, and transitions between different chiral phases~\cite{doi:10.1143/JPSJ.51.2478,0953-8984-2-47-015,doi:10.1063/1.4929447,NAGAMIYA1982385}. This work lays out the groundwork for a future fully relativistic DFT-DLM investigation of Mn$_3$A.

\begin{acknowledgments}
The authors gratefully acknowledge fruitful discussions with Christopher E.\ Patrick. The work was supported by EPSRC (UK) grants  EP/J06750/1 and EP/M028941/1.
\end{acknowledgments}

\appendix
\section{Electronic origin of the local internal fields and the direct correlation function}
\label{S2APP}

Multiple Scattering Theory (MST) based on the Korringa-Kohn-Rostoker (KKR) method~\cite{KORRINGA1947392,PhysRev.94.1111} is used to solve the Kohn-Sham equations in our DFT calculations.
The central quantities of our theory, i.e.\ the internal fields $\{\textbf{h}^\text{int}_n\}$ and the direct correlation function $\tilde{S}^{(2)}_{ss'}(\textbf{q})$, are directly given from characteristic objects of such a formalism, namely the $t$-matrix $\underline{t}_n(\hat{e}_n)$ describing a scattering event at site $n$ of an atom, which can contain a local moment oriented along $\hat{e}_n$, and the scattering path operator $\underline{\tau}_{nn'}$ describing and connecting a collection of scatterers~\cite{PhysRevB.5.2382}.
Underlines are used to indicate matrices in their angular momentum form.
The averages over $\{\hat{e}_n\}$ are conveniently performed directly from KKR-MST calculations~\cite{PhysRevB.5.2382,0305-4608-10-4-017} by constructing an effective medium that produces the averaged behavior of the scattering effects using the Coherent Potential Approximation (CPA)~\cite{PhysRev.156.809,PhysRevB.5.2382}.
$\textbf{h}^\text{int}_n$ is expressed as~\cite{PhysRevB.74.144411}
\begin{equation}
\textbf{h}^\text{int}_n=\frac{\text{Im}}{\pi}\int\text{d}\hat{e}_n\,\frac{\partial P_n(\hat{e}_n)}{\partial \textbf{m}_n}\int_{-\infty}^{\infty}\text{d}E f(E)\log\det\underline{D}^{-1}_n(\hat{e}_n),
\label{EQAPP-1}
\end{equation}
where $P_n(\hat{e}_n)$ is given in Eq.\ (\ref{EQ2-Theo}), $f(E)$ is the Fermi-Dirac function, and
\begin{equation}
\underline{D}_{n_0}(\hat{e}_{n_0})
=\left[
\underline{1}+\Big(\underline{t}^{-1}_{n}(\hat{e}_{n})-\underline{t}^{-1}_{c,n}\Big)
\underline{\tau}_{c,nn}\right]^{-1}
\label{EQAPP-2}
\end{equation}
is known as the impurity matrix, which is provided by the construction of the effective medium described by $\underline{t}_{c,n}(\hat{e}_n)$ and $\underline{\tau}_{c,nn}$. 
$\tilde{S}^{(2)}_{ss'}(\textbf{q})$ is obtained by applying the lattice Fourier transform to the derivative of Eq.\ (\ref{EQAPP-1}) with respect to $\textbf{m}_n$ in the paramagnetic limit~\cite{0305-4608-15-6-018,PhysRevLett.82.5369}. In this work we use a Fourier transform introduced in Eq.\ (\ref{EQ7-Wave3}) in order to handle complex multi-atom per unit cell structures. This gives
\begin{equation}
 \tilde{S}^{(2)}_{ss'}(\textbf{q})=
-\frac{\text{Im}}{\pi}
\int_{-\infty}^{\infty}\text{d}Ef(E)\text{Tr}(\underline{X}_{s}^{+}-\underline{X}_{s}^{-})\underline{\mathcal{I}}{ss'}(\textbf{q}),
\label{EQAPP-3}
\end{equation}
which is expressed in terms of a convolution integral
\begin{equation}
\begin{split}
\underline{\mathcal{I}}_{ss'}(\textbf{q})= &
\sum_{s''}\frac{1}{V_\text{BZ}}\int\text{d}\textbf{k} 
\tilde{\underline{\tau}}_{c,ss''}(\textbf{k}+\textbf{q})\tilde{\underline{\Lambda}}_{s''s'}(\textbf{q})\tilde{\underline{\tau}}_{c,s''s}(\textbf{k}) \\
 & -\underline{\tau}_{c,ss}\tilde{\underline{\Lambda}}_{ss}(\textbf{q})\underline{\tau}_{c,ss},
\label{EQAPP-4}
\end{split}
\end{equation}
where
$\underline{X}_{n}^{\pm}=[(\underline{t}_{n\pm}^{-1}-\underline{t}_{c,n}^{-1})+\underline{\tau}_{c,nn}]^{-1}$, in which $^\pm$ stand for up and down directions in a local spin frame of reference where $\underline{t}_{n}$ is diagonal, 
$\text{Tr}$ traces over angular momentum numbers,
$V_\text{BZ}$ is the Brillouin zone volume,
and $\tilde{\underline{\Lambda}}_{ss'}(\textbf{q})=\frac{1}{2}\left(\underline{X}_{s}^--\underline{X}_{s}^+\right)-\underline{X}_{s}^+\underline{\mathcal{I}}_{ss'}(\textbf{q})\underline{X}_{s}^-$.
For further background on the electronic structure part of the problem we refer the reader to references \cite{0305-4608-15-6-018} and \cite{PhysRevLett.82.5369}.

\bibliography{./bibliography.bib}
\end{document}